\documentclass{aa}  

\usepackage{graphicx}	
\usepackage{amsmath}	
\usepackage{float}
\usepackage{color}
\usepackage{soul}
\usepackage{caption}
\usepackage{subfig}
\usepackage{txfonts}

\newcommand{\Msun}{M$_{\odot}$}
\newcommand{\mycomment}[1]{}
\usepackage{multicol}
\usepackage{hyperref}
\begin{document} 

\title{Reconstructing post-common envelope white dwarf+main sequence binary histories through inverse population synthesis techniques}

   \subtitle{A Case Study of ZTF Eclipsing Binaries}

   \author{S. Torres
          \inst{1,2}\fnmsep\thanks{Email;
    santiago.torres@upc.edu}
          \and
          M. Gili\inst{1}
           \and
          A. Rebassa-Mansergas\inst{1,2}
          \and
          A. Santos-Garc\'{\i}a\inst{1}
                    \and
          A. J. Brown\inst{1}
            \and
          S. G. Parsons\inst{3}
          }

\institute{Departament de F\'\i sica, 
           Universitat Polit\`ecnica de Catalunya, 
           c/Esteve Terrades 5, 
           08860 Castelldefels, 
           Spain
           \and
           Institute d'Estudis Espacials de Catalunya, Esteve Terradas, 1, Edifici RDIT, Campus PMT-UPC, 08860 Castelldefels, Spain
           \and
           Department of Physics and Astronomy, University of Sheffield, Sheffield, S3 7RH, UK}

 \date{\today}
\titlerunning{WDMS inverse population synthesis techniques}
\authorrunning{Torres et al.}

\offprints{S. Torres}
 
  \abstract
   {The evolution of binary stellar systems involves a wide range of physical processes, many of which are not yet well understood. This is particularly true for close binary systems formed of a white dwarf and a main-sequence star. For instance, characterizing certain mass transfer episodes that may lead to a common-envelope phase and its subsequent evolution is still an open problem. Fortunately, the observational capabilities of current surveys, coupled with the feasibility of population synthesis models, enable us to reconstruct the past history of these systems, shedding light on their evolution and theoretical modeling.}
   {We aim to build a general-purpose algorithm based on inverse population synthesis techniques, able to reconstruct the past history of binary systems, particularly those involving a white dwarf and a main sequence star. This algorithm will be applied to a sample of eclipsing  binaries,  aiming to ascertain their progenitors and past histories.  Additionally, the resulting input space parameters will be analyzed, with a specific focus on the common-envelope phase.}
   {With the help of a consolidated population synthesis code, \texttt{MRBIN}, we developed an algorithm able to find the progenitor parameters of a given evolved binary system. The performance of the algorithm was tested on a set of synthetic binary systems. Once validated, it was applied to a sample 30 white dwarf plus main-sequence eclipsing binaries observed by the Zwicky Transient Facility survey. }
   {We determined the input space parameters of the progenitors for the 30 eclipsing binary systems to which the algorithm was applied. These parameters included the initial primary and secondary masses, the orbital separation and eccentricity, the common-envelope efficiency ($\alpha_{\rm CE}$), and the age at which the system was formed. Furthermore, the analysis of the global properties revealed some important features: a mild anticorrelation between the common-envelope efficiency parameter and the secondary mass, the absence of a universal value of $\alpha_{\rm CE}$ along with no need for internal energy, although in the low-mass regime, the high values of $\alpha_{\rm CE}$ suggest a possible contribution, and an initial thermalized eccentricity distribution. }
   {Although a strong degeneracy among the input parameters exists in the reconstruction of post-common envelope binary systems, the high accuracy obtained for the eclipsing-binary systems analyzed here has allowed our algorithm to make a reasonable determination of the initial parameters without the need to include external constraints. The global properties found here so far, can be substantially improved when analyzing a future volume-complete sample.}


   \maketitle
%

\section{Introduction}

A significant fraction of stellar systems in our Galaxy are formed by binary stars \citep[e.g.][]{Moe2017,Niu2021,Torres2022}. The physical processes involved in the evolution of such systems are more varied and, at the same time, often more complex than those describing the evolution of their single counterparts. Angular momentum transfer, magnetic braking, gravitational wave emission, rejuvenation and ageing, orbital synchronization, and, especially, mass transfer episodes, are some examples of such physical processes characteristic of binary evolution.

Unfortunately, many aspects of these processes remain poorly understood. Precise observational data coupled with comprehensive statistical analyses are essential to constrain certain theoretical models. Of particular interest are those binary systems formed by at least a white dwarf. These objects represent the most common remnant among stellar evolution and their physical properties are reasonably well understood from a theoretical point of view \citep[see][for recent reviews]{Althaus2010,Isern2022}. This understanding enables the derivation of precise evolutionary cooling models, thereby establishing white dwarfs as reliable cosmochronometers \citep{Fontaine2001}. The photometric magnitudes of white dwarfs are thus correlated with their cooling age. Hence, with an appropriate initial-to-final mass relationship we can derive the mass of the progenitor stars, their lifetime in the main-sequence, and ultimately, the total age of the system. This information, when coupled with other properties of the non-evolved companion, such as their metallicity, kinematics, or activity, can lead to important age-relation estimates \citep[e.g][]{Rebassa2021,Raddi2022,Rebassa2023}.

However, during a mass transfer episode, the inference of the initial progenitor's parameters can be especially challenging. This is particularly true when dynamically unstable mass transfer occurs, thus leading to mass overflow through the Roche lobe. This generally results in a common envelope phase through which the binary separation decreases dramatically. When formed of a white dwarf and a main-sequence star, the resulting system typically exhibits a short orbital period ranging from a few hours to a few days \citep{Rebassa2008, Nebot2011}. These systems are of special interest as precursors of cataclysmic variables \citep[e.g.][]{Pala2022}, super-soft X-ray sources \citep[e.g.][]{Parsons2015} and double degenerates \citep[e.g.][]{Napiwotzki2020}. All these exotic objects may later result in a type Ia supernovae \citep[e.g.][]{Wang+Han2012}.

Several attempts to reconstruct the progenitors of these kinds of systems, i.e., the initial conditions of the parent stars, can be found in the literature \citep[e.g.][and references therein]{Nelemans2005,Zorotovic2010,Davis2011,Hernandez2021}. However, given the  irreversible nature of many stellar processes (particularly the common-envelope phase), reversing this process to precisely reconstruct the complete past history solely from final observable parameters is a tough task. An approach to avoid these issues is to adopt fixed values for certain parameters or relying on a grid of predefined models. However, this limits the exploration of the initial parameter space, which, in our opinion, is one of the major drawbacks of the methodologies used so far.

Detailed binary evolution codes, such \texttt{MESA} \citep{Paxton2015} or \texttt{BINSTAR}  \citep{Siess2013}, allow for precise tracking of the physical processes governing the evolution of binary systems. However, these tools are generally time-consuming and can only simulate stellar evolution forward in time based on initial conditions.  On the other hand, binary population synthesis allows for faster, although less precise, simulations of a sample of binary stars from a given set of conditions. Hence, techniques to effectively identify the progenitors of a specific set of binary stars using population synthesis are required. Recent advances, such as the \texttt{dart\_board} code \citep{Andrews2018}, integrate Markov Chain Monte Carlo methods with rapid binary evolution models, improving efficiency and allowing for flexible modeling of both populations and individual binaries, while naturally incorporating observational uncertainties and additional constraints.

In this paper, we aim to develop an algorithm based on a population synthesis code capable of identifying the initial parameters that, through binary evolution, result in a set of final parameters consistent with the observable ones. No restrictions are imposed on the initial parameters, allowing us to identify possible correlations among these parameters. Time efficiency and the ability to be applied to a wide range of binary systems are also two desirable characteristics of our algorithm. Finally, the algorithm will be applied to a set 30 white dwarf plus main-sequence (WDMS) eclipsing binaries identified by the Zwicky Transient Facility survey \citep{Brown2023}. The accurate estimation of stellar parameters in these systems represents a clear advantage over previously used samples subjected to substantially larger observational uncertainties and will result in a restricted sample of compatible progenitors. Consequently, this restriction will refine the conditions of theoretical models and shed light on the evolution of these systems.

\section{Population synthesis code}
\label{s:popy}

In this work, the \texttt{MRBIN} population synthesis code is used for the simulation of binary systems. It was initially based on the binary stellar evolution code (\texttt{BSE}) developed by \cite{Hurley2002} but incorporated several updates provided by \cite{Camacho2014}, \cite{Cojocaru2017} and \cite{Canals2018}. The code is a multipurpose binary population synthesis simulator, but in particular it has been widely used for the study of the white dwarf population in binary systems \citep[see, for instance,][and references therein]{Torres2022}. 

Our code belongs to a family of synthetic stellar evolution codes (other examples include,  \texttt{binary\_c-python} \citet{Hendriks2023}; \texttt{StarTrack} \citet{Belczynski2008}; \texttt{SeBa} \citet{Toonen2012}, and references therein). These algorithms stand out for being relatively fast, as they rely on a set of adjustments of analytical formulas that replace precise and highly time-consuming calculations such as detailed stellar-structure and binary-evolution codes, e.g. \texttt{MESA} \citep{Paxton2015} or \texttt{BINSTAR}  \citep{Siess2013}. Despite this, the error introduced in the approximation does not exceed 5\% and they also enable a broader parameter analysis space compared to the more detailed evolution code counterparts \citep[see][for a recent and detailed analysis]{Fragos2023}.

Consequently, the use of \texttt{MRBIN} is justified as an initial proxy for the progenitor's solution. A detailed analysis of each system using a binary evolution code is beyond the scope of the present study and deferred to future work.

Several input parameters are needed to be considered for the  evolution of a binary system. Among them, the most relevant are the mass of the primary star (the, initially, more massive main sequence star), the mass  of the secondary star, the eccentricity, the separation and orbital period, the efficiency of the common-envelope phase, the metallicity, and the age when the system was formed (in our analysis we assumed that the formation of both stars are coeval).  Additionally, the code incorporates a set of prescriptions that account for other aspects of binary evolution such as tidal evolution, stellar winds, gravitational radiation, magnetic braking and angular momentum losses. All these processes introduce an additional set of parameters as described in \citet{Hurley2002}, which are kept constant throughout our simulations. 

For the treatment of the common-envelope phase we used the $\alpha$ prescription \citep{Webbink1984,Tout1997}. This formalism, essentially based on the energy conservation principle, introduces the $\alpha_{\rm CE}$ parameter as a measure of the efficiency of the orbital energy to unbind the common envelope of the system:
\begin{equation}
E_{\rm bin}=\alpha_{\rm CE}\Delta E_{\rm orb},
\end{equation}
where, $E_{\rm bin}$ is the binding energy of the envelope, and $\Delta E_{\rm orb}$ is the change in orbital energy. An approximation in a compact form for the binding energy is commonly employed:
\begin{equation}
E_{\rm bin}=-\frac{GM_{\rm donor}M_{\rm env}}{\lambda R_1},
\end{equation}
where $M_{\rm donor}$, $M_{\rm env}$, and $R_1$ are the total mass, the envelope mass and the radius, respectively, of the primary star at the beginning of the common-envelope phase, and $\lambda$ is the binding energy parameter \citep{deKool1987,Hurley2002}.  This parameter accounts for the approximation of the previous equation with respect to the exact value of the gravitational binding energy. The $\lambda$ parameter is approximated by a series of values that mainly depend on the type of star and its envelope mass \citep[see][and references therein]{Claeys2014}. In addition to the binding energy, we can take into account the internal energy, mainly due to the hydrogen recombination energy. However, it has been estimated that this internal energy is negligible in the case of post-common envelope WDMS binaries \citep{Zorotovic2010,Camacho2014}, as the systems in our study sample. Consequently, we have adopted a null internal energy contribution in our model.

Finally, our code incorporates some updates with respect to BSE version. For instance, we used the \texttt{PARSEC} evolutionary tracks \citep{Bressan2012}, providing an updated version of the mass-radius and, thus, effective temperature-luminosity relation for the main sequence stars. Analogously, the white dwarf evolution is followed according to La Plata cooling tracks \citep{Althaus2015,Althaus2021,Camisassa2016,Camisassa2017,Camisassa2019}  and the corresponding photometry by using Koester's \citep{Koester2010} hydrogen-rich atmospheric models. The set of La Plata  models incorporate the most updated physical processes in white dwarf evolution, such as crystallization, phase separation and sedimentation of major species such as Ne22,  as well as covering the full range of chemical core compositions (helium, carbon-oxygen, and oxygen-neon).

\section{Methodology: inverse population synthesis}
\label{s:inpo}

Reconstructing the past history of post-common envelope WDMS binaries involves determining the parameters of their progenitors: their masses, separations, eccentricities, ages, and ultimately, a set of initial parameters. In the following sections, we outline the strategy presented in our work, which is based on the population synthesis code, \texttt{MRBIN}, described in the previous section.

\subsection{The inverse population synthesis algorithm}
\label{ss:inpoal}

\begin{figure}
	\includegraphics[width=\columnwidth,trim=90 250 290 50, clip]{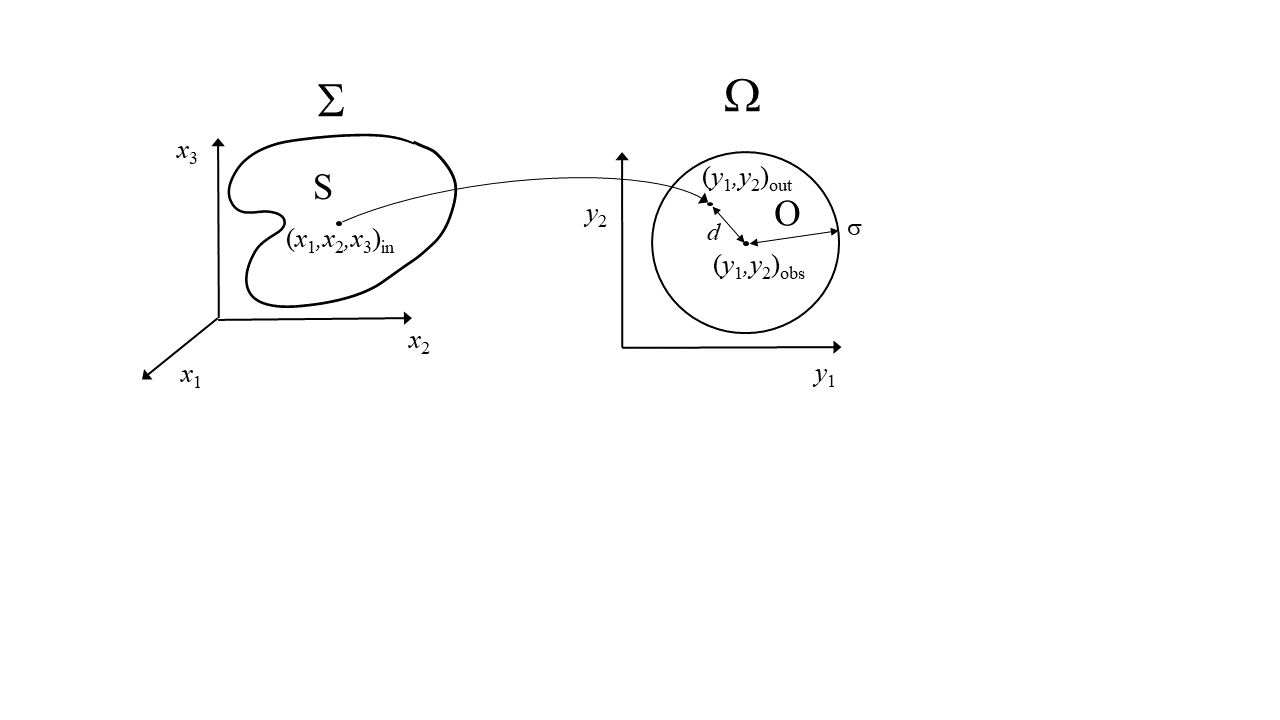}
    \caption{Schematic representation of the problem. The space that contains all possible initial values ($\Sigma$) is projected, after binary evolution, onto the output space ($\Omega$). Only those points within $S$ lead to a final set of parameters, called $O$, that are at a distance $d<\sigma$ from a certain set of {\sl observed} parameters. See text for details.}
    \label{f:space}
\end{figure}

In Figure~\ref{f:space} we show a schematic representation of the problem. $\Sigma$ is the space containing all possible initial values, while $\Omega$ represents the space containing all possible output values after binary evolution. Our goal is to find the subset $S\subset\Sigma$, such that a system, $\vec{x}_{\rm in}=(x_1,x_2,\ldots,x_n)_{\rm in}$ that  belongs to $S$, after binary evolution leads to the final set of parameters $\vec{y}_{\rm out}=(y_1,y_2,\ldots,y_m)_{\rm out}$, which is at a distance $d$ smaller than a certain $\sigma$ from the set of {\sl observed} parameters $\vec{y}_{\rm obs}=(y_1,y_2,\ldots,y_m)_{\rm obs}$. As an initial assumption we can adopt a Euclidean normalized distance between the output point and the observed point:

\begin{equation}
d=\sqrt{\left(\frac{y_{1,{\rm out}}-y_{1,{\rm obs}}}{y_{1,{\rm obs}}}\right)^2+\left(\frac{y_{2,{\rm out}}-y_{2,{\rm obs}}}{y_{2,{\rm obs}}}\right)^2+\ldots+\left(\frac{y_{m,{\rm out}}-y_{m,{\rm obs}}}{y_{m,{\rm obs}}}\right)^2}
\label{eq:distance}
\end{equation}

The goal of the algorithm is to find $S$ in an efficient and computationally low-cost way. In order to do that the algorithm performs a random walk in the initial space ($\Sigma$) to identify all possible combinations of parameters that lead to a desired result in the observable space ($\Omega$). That is,  by applying random displacements, the algorithm explores a set of initial parameters which are evolved using the \texttt{MRBIN} population synthesis code into a set of output parameters. Next, we calculate the distance (Eq.\ref{eq:distance}) of this output point to a given point in the observable space. This distance is compared with an arbitrary $\sigma$, which represents the allowed error threshold (e.g., $1\%$, $5\%$, $10\%$, etc.). If the result is within the allowed error range (i.e.,  $d\leq \sigma$), the set of initial parameters is considered valid and saved. Otherwise, the point is discarded and the algorithm returns to the previous point. Then, a new set of random displacements is applied and the process is repeated.  
 
To maximize the efficiency of the algorithm (that is, to find the border that delimits the solution space $S$ in the minimum number of steps), adaptive steps between points can be used. This means that if the initial generated point is considered valid, the next step taken from that point will be larger than the previous one. However, if the new point exceeds the maximum error allowed, the algorithm will return to the previous point and generate another one using a smaller step size. In our case, we adopt a scaling factor, $\epsilon$ of 1.25; that is, we increase or decrease each new random displacement by 25\%.

\begin{figure}
	\includegraphics[width=\columnwidth,trim=10 240 380 20, clip]{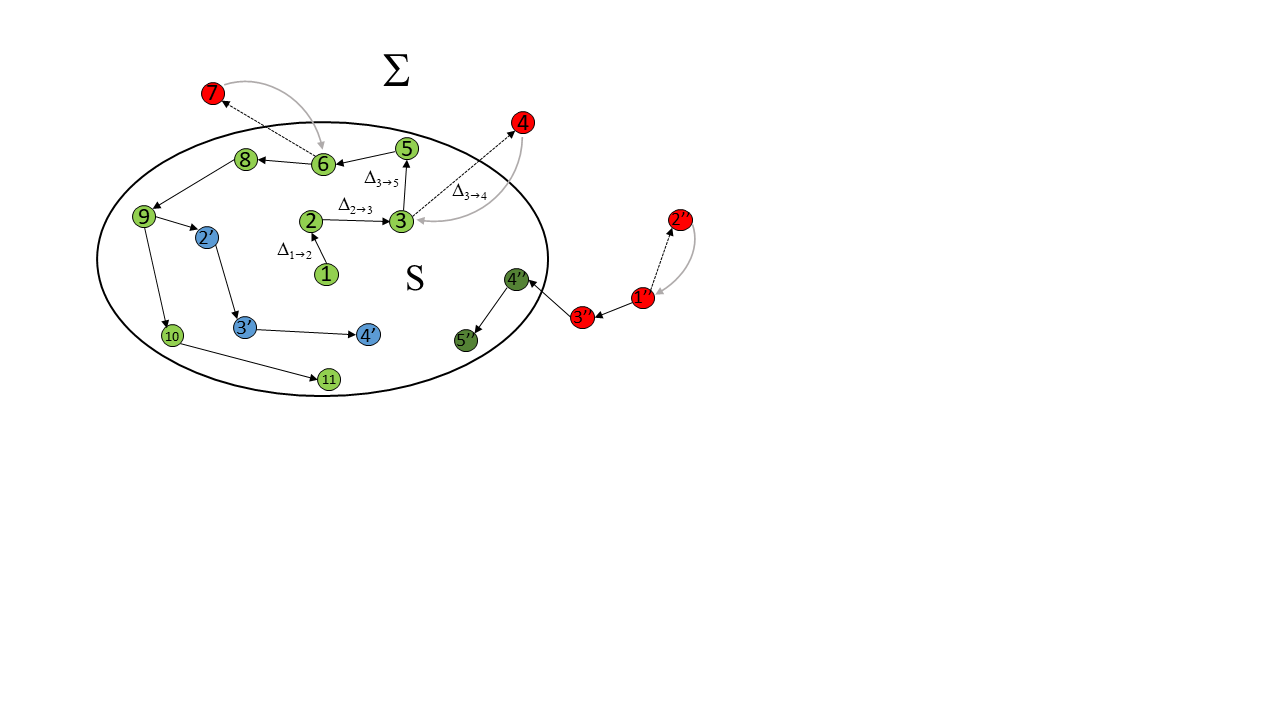}
    \caption{Schematic representation of our adaptive random walk algorithm. See text for details.}
    \label{f:random}
\end{figure}

In Figure~\ref{f:random} we plot an example of the random walk process in a 2-dimensional space. It starts from point 1 (green circle) inside the solution space $S$. It makes a first step of length $\Delta_{1\rightarrow 2}$ to point 2, i.e., a random displacement in each of the $n$-dimension of the input space $\Sigma$. The new point is evolved and checked to be valid (i.e., $d\leq \sigma$), so that the algorithm continues from it. The next step is now increased by the scaling  factor $\epsilon=1.25$, so that $\Delta_{2\rightarrow 3}$ is on average a 25\% larger than before.  The algorithm falls to point 3, which is also valid. The next step is even larger, but in this case point 4 is invalid (i.e.,  $d>\sigma$), so the algorithm has to return to point 3 and do a new step from there, now a 25\% smaller. After two valid points (5 and 6), the same happens with point 7, implying that the algorithm goes back to point 6 and continues in another direction with a smaller step. Points 8, 9 and 10 are consecutively valid, so that larger and larger steps are taken.

To minimize the time required for the algorithm to explore the solution space, we initiate a new random walk from any arbitrary solution point that has been found thus far. For instance, point 9 is adopted as the new starting point for the random walk, with points 2', 3', and 4' marked in blue. Since this new random walk operates independently from the previous one, both can be executed in parallel. Moreover, since this process can be applied iteratively, it leads to a significant reduction in the overall computation time.

Finally, if the initial guess does not belong to the solution space, $S$, a similar adaptive random walk process is applied until we locate a solution within $S$. In our example of Fig.~\ref{f:random} that would be the case when starting with point 1'' (red circle). If the new point, 2'', is at a distance $d$ larger than the previous one, we return to the original point since we find a new one, point 3'', whose distance is smaller. This process is repeated until we finally find a point, 4'', that belongs to the solution space $S$.  It then continues as a random walk as previously described.

\subsection{Validation of the algorithm}
\label{ss:valalg}

We have tested the feasibility of our algorithm with the assistance of our population synthesis code, \texttt{MRBIN}. Since, for synthetic systems, we know both the initial and final values, we can thus verify the capability of our algorithm to find solutions given a set of final values that would play the role of observed values. Different combinations of input and output parameters have been tested, as well as different error thresholds ($\sigma$).

In all cases, we have been able to verify that the algorithm based on adaptive random walks can find a stable set of solutions. That is, the distribution of the initial parameters is sufficiently smooth, and they are independent of the algorithm's starting point.  This is accomplished as long as the number of generated points is sufficiently high (on the order of $10^6$), and a minimum of 100 random walks are generated.

For illustrative purposes we have chosen here three representative cases that correspond to WDMS systems with final periods of the order of a few hours, ten hours, and days ($P=1.16,\,15.50,$ and $41.73\,$h, respectively). The output space consists of five variables (these five variables are the ones that we can get observationally, see Section \ref{s:obsa}): the masses and effective temperatures of the WDMS, and the orbital period of the system. The free parameters correspond to six variables of the input space: the initial masses of the stars, the initial period and eccentricity, the common envelope efficiency parameter and the age of the system. The metallicity has been fixed to the solar value, and for the three systems shown here we adopted  $\alpha_{\rm CE}=0.3$.

\begin{figure*}
\begin{multicols}{2}
    \includegraphics[width=0.85\columnwidth,trim=-90 -5 10 10, clip]{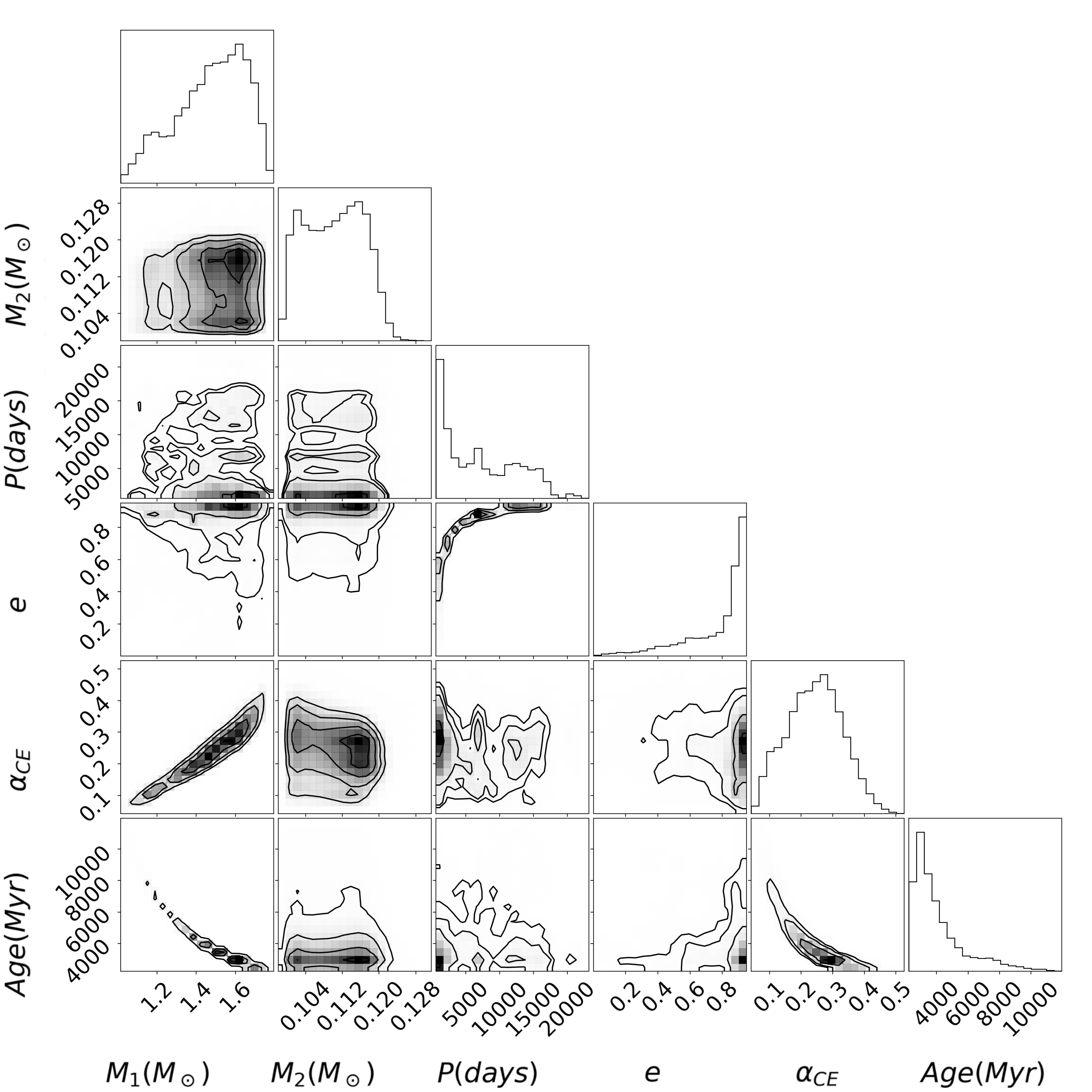}\par 
    \includegraphics[width=1.15\columnwidth,trim=80 15 -20 10, clip]{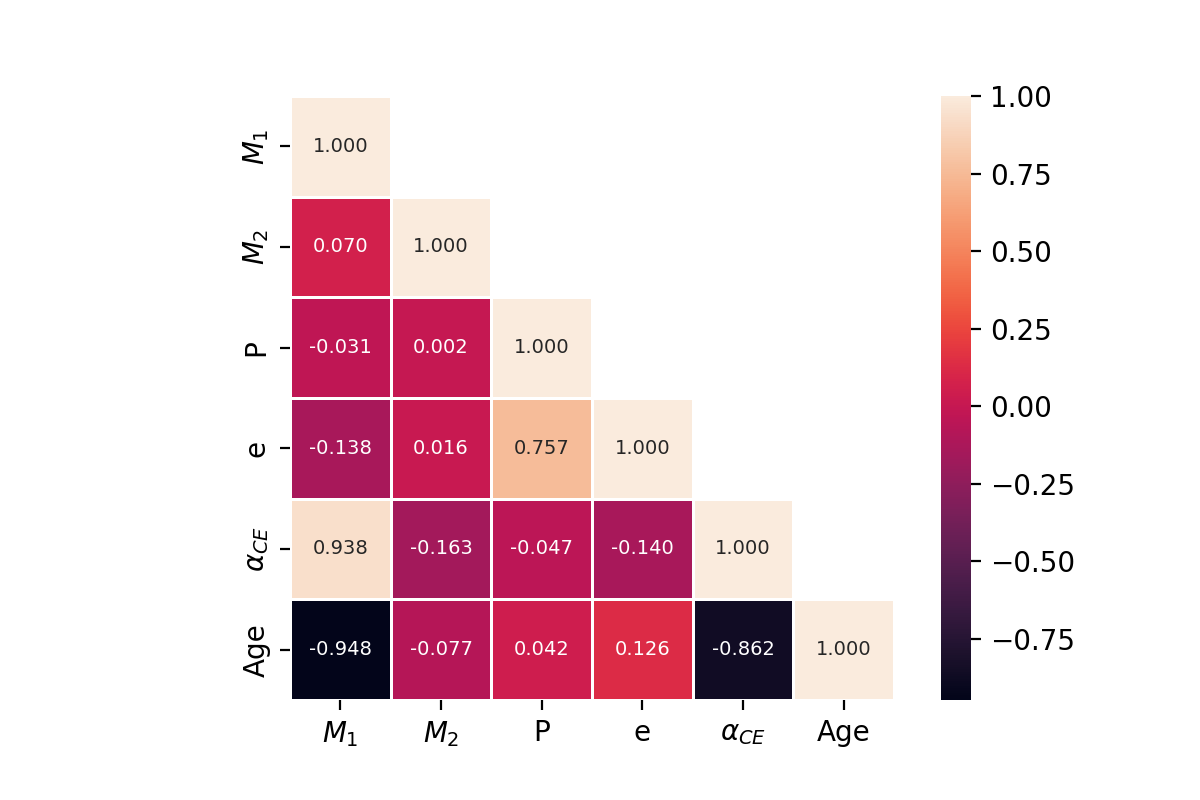}\par 
    \end{multicols}
\begin{multicols}{2}
    \includegraphics[width=0.85\columnwidth,trim=-90 -5 10 10, clip]{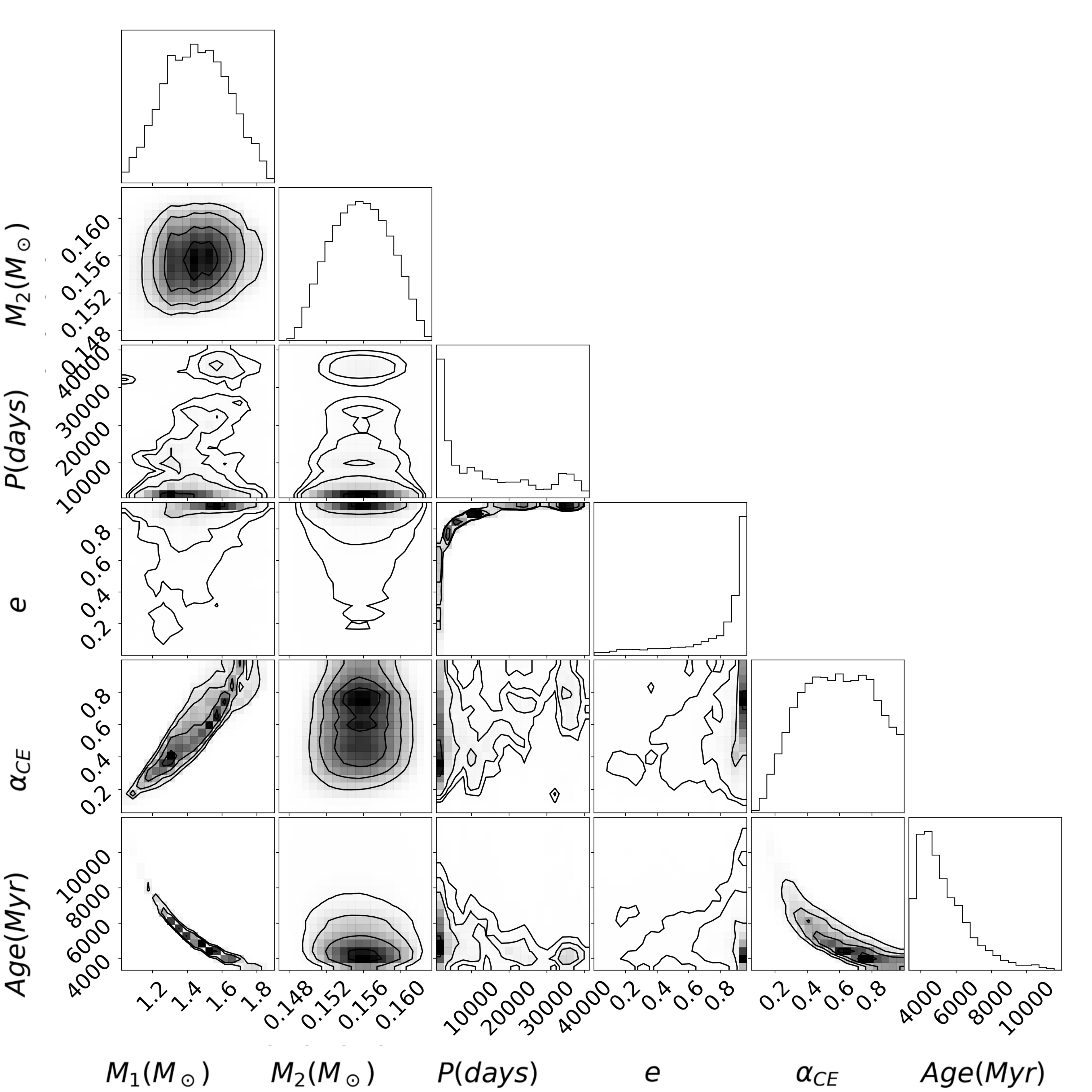}\par
    \includegraphics[width=1.15\columnwidth,trim=80 15 -20 10, clip]{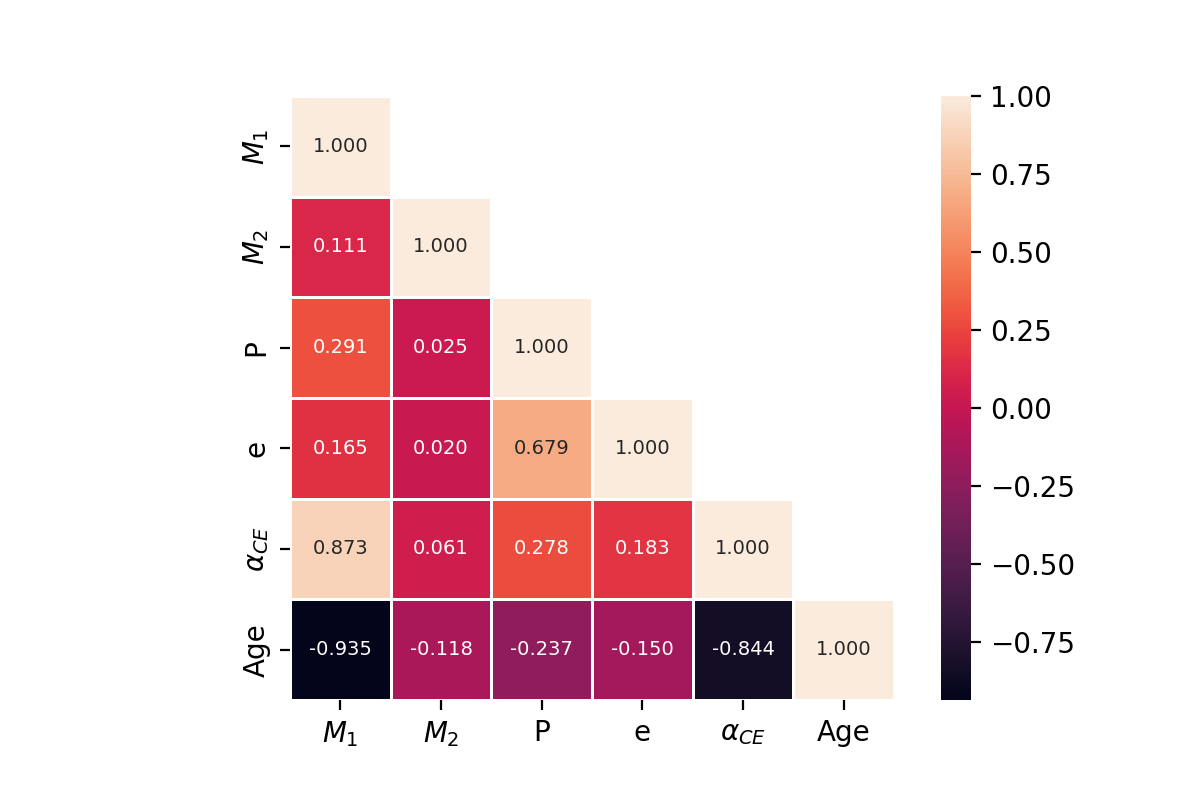}\par
\end{multicols}
\begin{multicols}{2}
    \includegraphics[width=0.85\columnwidth,trim=-90 -5 10 10, clip]{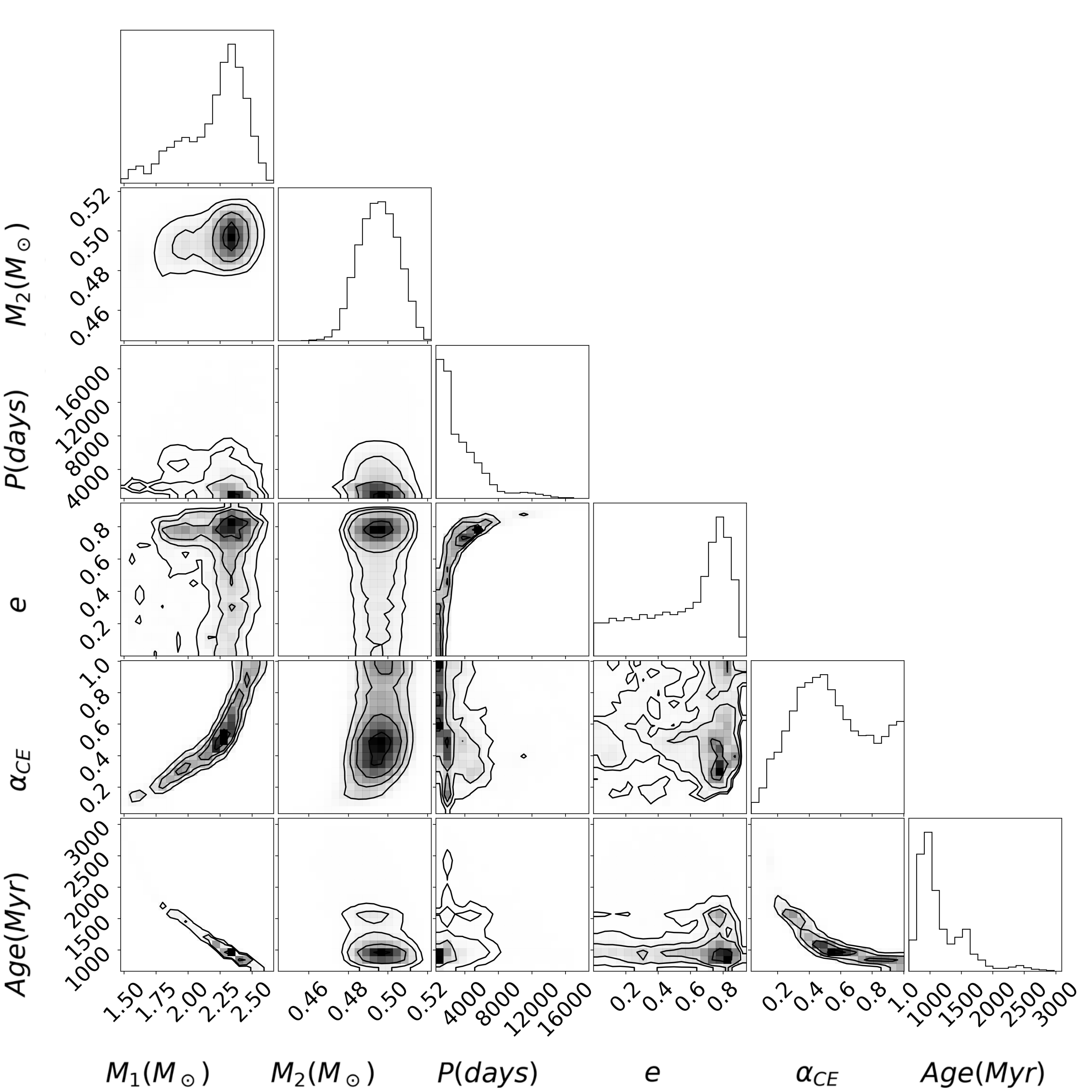}\par
    \includegraphics[width=1.15\columnwidth,trim=80 15 -20 10, clip]{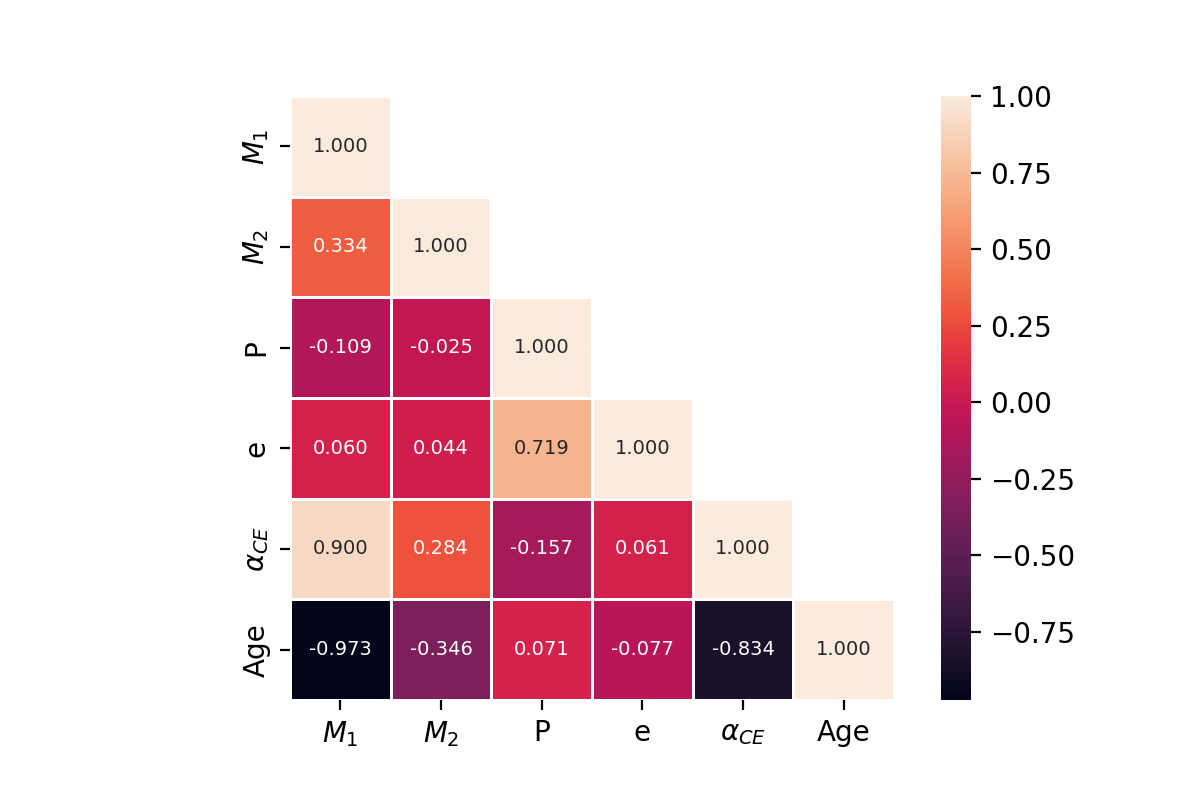}\par
\end{multicols}
\caption{Inverse population synthesis validation test: corner plots (left panels) and correlation matrices (right panels) corresponding  to the three synthetic binary systems with final periods $P=1.16$, $P=15.5$, and $P=41.73$ hours (top, middle and bottom panels, respectively). Corner plots represent the set of solutions found by our algorithm within a 1\% error threshold. See text for details.}
    \label{f:val}
\end{figure*}

Results are displayed in Figure \ref{f:val} using a corner plot, which represents the marginal distribution of each of our considered input parameters (top histograms) and density maps of possible projections between any two parameters (rest of sub-panels). Furthermore, correlation matrices are also  generated in order to identify any potential relationships between parameters. Each term of the matrix, Pearson correlation coefficient, is calculated as:
\begin{equation}
    r = \frac{{\sum((X - \overline{X})\cdot(Y - \overline{Y}))}}{{\sqrt{{\sum(X - \overline{X})^2}} \cdot \sqrt{{\sum(Y - \overline{Y})^2}}}}
\end{equation}
where $X$ and $Y$ represent the respective values of two parameters, while $\overline{X}$ and $\overline{Y}$ denote the means of the $X$ and $Y$ variables, respectively.  We recall here that correlation coefficients range between -1 and 1 in the off-diagonal elements of a correlation matrix. A value of -1 indicates a perfect negative linear dependence, +1 indicates a perfect positive linear dependence, and 0 indicates no linear dependence at all.

\begin{table*}[h!]
    \centering
    \caption{Validation test: synthetic input parameters for the
three systems analyzed here (Systems 1 to 3, rows), along with the corresponding values for the optimal solution (Joint prob., rows) for the three threshold distances: 0.1\%, 1\%, and 10\%, respectively.}
    \label{tab:validation}
    \begin{tabular}{cccccccc}
        \hline
         & $M_{1,{\rm in}}$ (M$_{\odot}$) & $M_{2,{\rm in}}$ (M$_{\odot}$) & $P_{\rm in}$ (days) & $e_{\rm in}$ & $\alpha_{\rm CE}$ & Age (Myr) & $d_{\rm min}$ (\%) \\
        \hline
        \hline
       System 1  & 1.477 & 0.103 & 3613.3 & 0.781 & 0.300 & 4496.6  & 0\\
     Joint prob. & 1.476 & 0.103 & 3552.1 & 0.778 & 0.300 & 4506.4  & 0.1\\
        Joint prob. & 1.603 & 0.151 & 3294.5 & 0.802 & 0.236 & 3795.0 & 1\\
      Joint prob. & 1.506 & 0.134 &	840.0 &	0.381	& 0.255	& 4219.6 & 10\\
        \hline
      System 2 & 1.201 & 0.156 & 3117.56 & 0.751 & 0.300 & 6976.77 & 0\\
       Joint prob.  & 1.203 & 0.156 & 3205.18 & 0.749 & 0.292 & 6937.81 & 0.1\\
      Joint prob. & 1.318 & 0.156 & 1068.40 & 0.349 & 0.443 & 6597.66 & 1\\
       Joint prob. & 1.446	& 0.157 &	7046.16 & 0.897 & 0.745 & 3961.19 & 10\\
        \hline
       System 3 & 1.926 & 0.493 & 2432.85 & 0.574 & 0.300 & 1508.59 & 0\\
         Joint prob.  & 1.922 & 0.493 & 2442.10 & 0.575 & 0.298 & 1511.57 & 0.1\\
         Joint prob.  & 1.947 & 0.493 & 3880.81 & 0.723 & 0.327 & 1503.77 & 1\\
         Joint prob.  & 2.345 & 0.500 &  1996.82 &	0.776 & 0.996 & 930.14 & 10\\
        \hline
    \end{tabular}
\end{table*}

The analysis of the results presented in Fig. \ref{f:val} reveals a similar behavior of the space parameter  regardless of the final period of the systems. In all cases, a smooth distribution of the input parameters is found, indicative that the algorithm has reached a stable set of solutions. Moreover, a positive linear dependence, $r\sim0.9$, is presented between the common envelope efficiency, $\alpha_{\rm CE}$, and the mass of the white dwarf progenitor, $M_1$. Similarly, a negative dependence is found between the age of the system and $M_1$ and $\alpha_{\rm CE}$. These dependencies reflect the fact that a specific final period can be achieved by a more massive primary progenitor in a shorter time when using a larger $\alpha_{\rm CE}$ parameter. Conversely, when the primary progenitor is less massive, a smaller $\alpha_{\rm CE}$ is required, but a longer age is needed.

The previously mentioned high correlation reflects a strong degeneracy in the solution space, $S$.  Different combinations of the initial parameters (see left panels of Fig. \ref{f:val}) reproduce a set of similar final parameters within a certain $\sigma$ threshold (1\% in the case of the systems represented in Fig. \ref{f:val}). Among the set of possible solutions, $\vec{x}_s\in S$,  we have adopted the most representative or optimal one, $\vec{x}_{\rm opt}$, as the most probable one. We estimated the optimal solution as that which maximizes the joint probability distribution,
\begin{equation}
\label{e:joint}
    p(\vec{x}_{\rm opt})=\max{\prod_i} f_{X_i}(x_i) \quad | \quad \vec{x}\in S,
\end{equation}
where $f_{X_i}(x_i)$ is the marginal distribution of the input parameter $X_i$. The fact that the probability is constrained to those points, $\vec{x}$, belonging to the solution space $S$, implicitly incorporates the possible correlations between parameters. Moreover, we can estimate the error in the optimal solution from those points, $\vec{x}_{\rm err}$, whose likelihood-based confidence region is within $1\sigma$ from the probability of the optimal solution $\vec{x}_{\rm opt}$, i.e., $p(\vec{x}_{\rm err})\ge p(\vec{x}_{\rm opt})/\sqrt{e}$.

In Table \ref{tab:validation}, we present the synthetic input parameters for the three systems analyzed here, along with the corresponding values for the optimal solution (the one that maximizes the joint probability, Eq.\ref{e:joint}) for the three threshold distances: 0.1\%, 1\%, and 10\%, respectively. The analysis of Table \ref{tab:validation} reveals that the input parameters found by our algorithm are either perfectly close or quite close to the synthetic ones for threshold distances of 0.1\% and 1\%, respectively. However, this is not the case when the threshold distance is increased to 10\%. In that situation, some of the input parameters, such as the period or eccentricity, clearly diverge from the synthetic ones. Even though, it is important to emphasize here that the solution found by our algorithm in that case is still completely valid. When enlarging the threshold distance, the space of possible solutions $S$ is also enlarged. In that situation, the synthetic input values used as a reference for our validation systems do not necessarily represent the ones that maximize the joint probability distribution. Given the high correlation between some of the input parameters, solutions with a higher probability than the one derived from the synthetic values can be found.

Nevertheless, some caveats and warnings regarding the method should be addressed. First, provided that a sufficiently large number of random walkers are generated, we can reasonably assume that the method recovers the entire solution space consistent with the observational constraints. Second, observational uncertainties must be small enough so that the corresponding solution space $S$ is also narrow. In this limited region, we assume that prior distributions (such as the initial mass function) can be approximated as approximately constant. Thus, the impact of including full priors would introduce only second-order corrections to the recovered parameter distributions.

 As long as these two conditions are met, we can approximate the optimal solution by maximizing the product of the marginal distributions, thereby minimizing the effects of possible correlations between parameters. While this approach has both advantages and limitations, one of our broader goals is to ultimately derive an estimate of the priors from the collective properties of the final set of solutions (see Section \ref{ss:global}).

In this sense, our method is not strictly a Markov Chain Monte Carlo method, as described in \cite{Andrews2018}, since it does not apply a probability criterion based on the posterior distribution to guide the random walks. Instead, our approach is closer to a density estimation technique, aiming to approximate the probability density function rather than merely sampling from it. In conclusion, the algorithm presented here for inverse population synthesis reconstruction can be considered as a feasible tool for exploring the entire space of initial variables and efficiently finding the set of solutions and their possible correlations, within a reasonable low computational cost.

\section{The observed sample}
\label{s:obsa}

In this work we analyze the sample presented in \cite{Brown2023}. In their work, they conducted a photometric follow-up of 43 eclipsing WDMS binaries detected by the Zwicky Transient Facility (ZTF) survey\footnote{\url{https://www.ztf.caltech.edu/}}, performing an extensive photometric campaign, and acquiring precise light curves for each object. Of the 43 systems that they followed-up, 9 do not have measured parameters because they harbour magnetic white dwarfs. In addition, the light-curve fits of 4 objects suggest the white dwarf companions to be brown dwarfs, so that the best fit secondary masses can only be regarded as upper limits. Taking this into account we will apply the algorithm to the remaining 30 systems. After fitting the best model to each observed system,  \cite{Brown2023} found a set of stellar parameters for each  component of the system. In Table \ref{tab:ztf} we show the most relevant stellar parameters obtained in that work and the corresponding observational errors. Among the parameters available in their study, we adopted the more representative ones, including the observed masses and effective temperatures of both components, as well as the orbital period of the system. Typical observational errors for these parameters are of the order of 5\%.

\begin{table*}[h!]
    
    \caption[Observed sample]{Stellar parameters obtained for 30 selected eclipsing binaries from \cite{Brown2023} and the metallicity value $Z$ derived in this work.}
    \label{tab:ztf}
    \begin{center}
 \begin{tabular}{lcccccc}
  \hline
    Target & $M_1$ (M$_{\odot}$) & $T_{\rm eff_1}$ (K) & $M_2$ (M$_{\odot}$) & $T_{\rm eff_2}$ (K) & $P$ (days) & $Z$\\
    \hline
    \hline
 ZTF J041016.82-083419.5 & $0.355_{-0.011}^{+0.015}$ & $14690_{-550}^{+560}$ & $0.123_{-0.008}^{+0.009}$ & $2840_{-110}^{+110}$ & 0.0811093 & 0.008 \\
 ZTF J051902.06+092526.4 & $0.391_{-0.029}^{+0.019}$ & $10750_{-580}^{+770}$ & $0.177_{-0.019}^{+0.014}$ & $2800_{-110}^{+140}$ & 0.0929131 & 0.026 \\  
  ZTF J052848.24+215629.0 & $0.787_{-0.025}^{+0.025}$ & $12100_{-630}^{+700}$ & $0.184_{-0.013}^{+0.014}$ & $3130_{-110}^{+110}$ & 0.2259952 & 0.008\\ 
  ZTF J053708.26-245014.6  & $0.397_{-0.007}^{0.009}$ & $16100_{-410}^{+440}$ & $0.204_{-0.011}^{+0.012}$ & $2970_{-100}^{+100}$ & 0.3277936 & 0.019  \\
  ZTF J061530.96+051041.8  & $0.560_{-0.011}^{+0.011}$ & $15220_{-510}^{+600}$ & $0.533_{-0.029}^{+0.030}$ & $3380_{-110}^{+110}$ & 0.3481742 & 0.017 \\ 
  ZTF J063808.71+091027.4 & $0.604_{-0.011}^{+0.013}$ & $22500_{-1000}^{+1200}$ & $0.410_{-0.022}^{+0.024}$ & $3320_{-110}^{+110}$ & 0.6576453& 0.029 \\ 
  ZTF J063954.70+191958.0 & $0.701_{-0.009}^{+0.011}$ & $15980_{-520}^{+520}$ & $0.210_{-0.011}^{+0.011}$ & $3200_{-100}^{+100}$ & 0.2593556 & 0.008 \\ 
  ZTF J064242.41+131427.6 & $0.633_{-0.008}^{+0.011}$ & $14560_{-500}^{+540}$ & $0.150_{-0.008}^{+0.008}$ & $3110_{-100}^{+100}$ & 0.1710542 & 0.006 \\ 
  ZTF J065103.70+145246.2 & $0.515_{-0.020}^{+0.019}$ & $13140_{-670}^{+560}$ & $0.242_{-0.019}^{+0.018}$ & $3170_{-110}^{+120}$ & 0.1677075 & 0.012\\ 
  ZTF J070458.08-020103.3  & $0.500_{-0.015}^{+0.012}$ & $9280_{-250}^{+230}$ & $0.344_{-0.020}^{+0.018}$ & $3300_{-100}^{+100}$ & 0.1413708 & 0.013 \\ 
  ZTF J071759.04+113630.2 & $0.528_{-0.017}^{+0.016}$ & $21110_{-750}^{+720}$ & $0.296_{-0.022}^{+0.020}$ & $3150_{-110}^{+120}$ & 0.4527638 & 0.018\\ 
  ZTF J071843.68-085232.1 & $0.794_{-0.018}^{+0.019}$ & $18940_{-880}^{+870}$ & $0.306_{-0.019}^{+0.020}$ & $3120_{-110}^{+110}$ & 0.2158113 &  0.025\\ 
  ZTF J080542.98-143036.3 & $0.393_{-0.013}^{+0.013}$ & $26500_{-9000}^{+1200}$ & $0.291_{-0.023}^{+0.020}$ & $3250_{-110}^{+120}$ & 0.1981669 & 0.011 \\
  ZTF J094826.35+253810.6 & $0.504_{-0.024}^{+0.026}$ & $11290_{-450}^{+480}$ & $0.16_{-0.014}^{+0.015}$ & $3120_{-120}^{+120}$ & 0.1418270 & 0.006 \\ 
  ZTF J102254.00-080327.3 & $0.605_{0.025}^{+0.027}$ & $8330_{-250}^{+260}$ & $0.405_{-0.029}^{+0.030}$ & $3170_{-110}^{+110}$ & 0.2314179 & 0.033\\ 
  ZTF J102653.47-101330.3 & $0.376_{-0.010}^{+0.012}$ & $19320_{-670}^{+710}$ & $0.105_{-0.006}^{+0.008}$ & $2840_{-110}^{+110}$ & 0.0929868 & 0.004\\ 
  ZTF J104906.96-175530.7 & $0.426_{-0.007}^{+0.010}$ & $13000_{-460}^{+440}$ & $0.198_{-0.010}^{+0.012}$ & $3170_{-110}^{+100}$ & 0.2447332 & 0.008 \\ 
  ZTF J122009.98+082155.0 & $0.580_{-0.018}^{+0.017}$ & $10170_{-260}^{+270}$ & $0.275_{-0.020}^{+0.019}$ & $3140_{-110}^{+110}$ & 1.2329254 & 0.018\\ 
  ZTF J125620.57+211725.8 & $0.479_{-0.009}^{+0.010}$ & $5073_{-79}^{+79}$ & $0.101_{-0.005}^{+0.005}$ & $2950_{-100}^{+100}$ &  0.5560572 & 0.002 \\ 
  ZTF J130228.34-003200.2 & $0.811_{-0.016}^{+0.021}$ & $11790_{-330}^{+400}$ & $0.179_{-0.010}^{+0.012}$ & $3030_{-100}^{+100}$ &  0.1661310 & 0.012\\ 
  ZTF J134151.70-062613.9 & $0.509_{-0.035}^{+0.038}$ & $58300_{-8700}^{+8400}$ & $0.126_{-0.009}^{+0.015}$ & $2800_{-220}^{+210}$ &  0.0969505 & 0.010 \\ 
  ZTF J140036.65+081447.4 & $0.563_{-0.008}^{+0.009}$ & $13340_{-610}^{+650}$ & $0.232_{-0.012}^{+0.012}$ & $2970_{-100}^{+100}$ &  0.2602766 & 0.026\\
  ZTF J140423.86+065557.7 & $0.736_{-0.015}^{+0.016}$ & $14980_{-460}^{+470}$ & $0.409_{-0.023}^{+0.023}$ & $3100_{-100}^{+100}$ &  0.1683096 & 0.039\\ 
  ZTF J140537.34+103919.0  & $0.404_{-0.008}^{+0.008}$ & $29900_{-9000}^{+1100}$ & $0.085_{-0.005}^{+0.005}$ & $3430_{-140}^{+130}$ &  0.2714122 & 0.014 \\ 
  ZTF J140702.57+211559.7 & $0.406_{-0.014}^{+0.018}$ & $10870_{-350}^{+350}$ & $0.263_{-0.016}^{+0.021}$ & $3160_{-110}^{+110}$ &  0.1432802 & 0.017\\ 
  ZTF J162644.18-101854.3 & $0.499_{-0.012}^{+0.015}$ & $36700_{-2700}^{+2700}$ & $0.212_{-0.011}^{+0.013}$ & $3180_{-110}^{+110}$ &  0.2530067 & 0.010 \\ 
  ZTF J163421.00-271321.7 & $0.436_{-0.054}^{+0.042}$ & $10680_{-630}^{+790}$ & $0.134_{-0.020}^{+0.016}$ & $2400_{-120}^{+130}$ &  0.0780396 & 0.026 \\ 
  ZTF J164441.18+243428.2 & $0.382_{-0.018}^{+0.020}$ & $13270_{-460}^{+520}$ & $0.103_{-0.009}^{+0.009}$ & $2500_{-110}^{+110}$ &  0.0801054 & 0.015 \\ 
  ZTF J180256.45-005458.3  & $0.458_{-0.021}^{+0.019}$ & $10770_{-500}^{+630}$ & $0.150_{-0.011}^{+0.010}$ & $3150_{-110}^{+110}$ &  0.2690033 & 0.005 \\
  ZTF J195456.71+101937.5 & $0.509_{-0.012}^{+0.015}$ & $21500_{-1100}^{+1000}$ & $0.449_{-0.026}^{+0.028}$ & $3480_{-110}^{+110}$ &  0.3102884 & 0.021 \\
\hline
    \end{tabular}
\end{center}
\end{table*}

To accurately reconstruct the past histories of these systems, it is also necessary to incorporate the metallicities of the stars, a parameter that is not provided by \cite{Brown2023}. We utilized the \texttt{PARSEC}  isochrones \citep{Bressan2012} for this purpose. Using the observed values of effective temperature and mass, we calculate the metallicities by performing multivariable interpolation on the corresponding isochrones. This procedure has been applied to our observed sample, obtaining a metallicity proxy value for all systems except for ZTF J140537.34+103919.0.   Due to its lowest mass companion \citep[reported as just above the hydrogen-burning limit,][]{Brown2023}, proper determination of its metallicity is not feasible. Therefore, a solar value of $Z=0.014$ has been adopted. The list of metallicities derived is presented in the  right column of Table \ref{tab:ztf}. An average value of $Z=0.0145$ with a dispersion of $\sigma_Z=0.003$ has been found, thus indicating the thin disk origin of the vast majority of this sample and the perfect agreement with metallicity estimates of WDMS binaries \citep{Rebassa2021}.

\section{Results}
\label{s:resu}

\subsection{Reconstructing individual binary systems}

Our algorithm has been applied to each of the 30 systems from \cite{Brown2023} presented in Table \ref{tab:ztf}. We remind the reader that the observed space parameter consists of the masses and effective temperatures of both stars, and the orbital period of the system.  For the input space parameter (that is, our free parameters) we adopted a six-dimensional space, consisting in the masses of the progenitor stars, the initial eccentricity and orbital period, the common envelope efficiency, and the age when the system was born. For each system we adopted the metallicity presented in Table \ref{tab:ztf} as a constant value along the simulation.

\begin{figure*}
\begin{multicols}{2}
    \includegraphics[width=0.95\columnwidth,trim=-90 -5 10 10, clip]{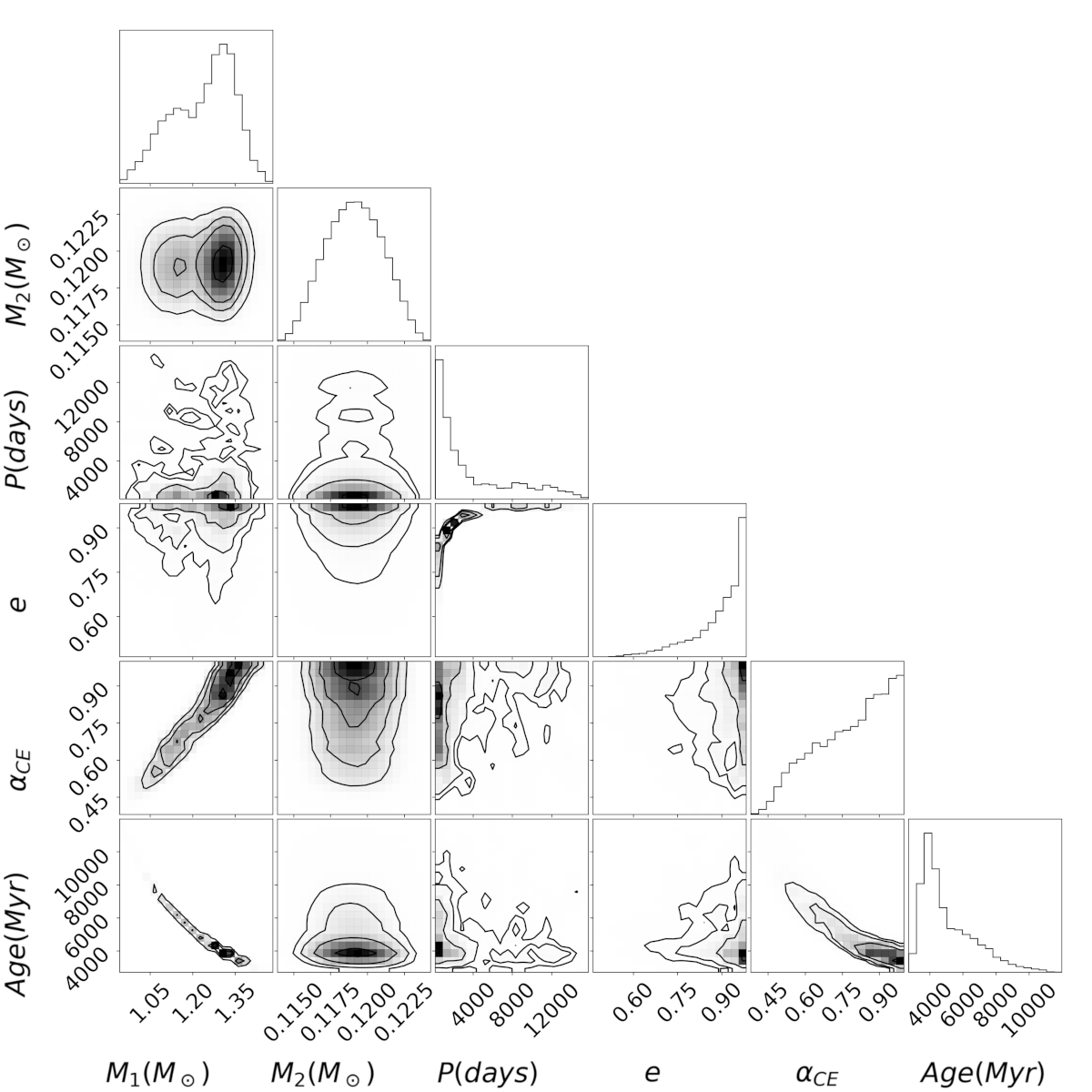}\par
    \includegraphics[width=1.15\columnwidth,trim=80 15 -20 10, clip]{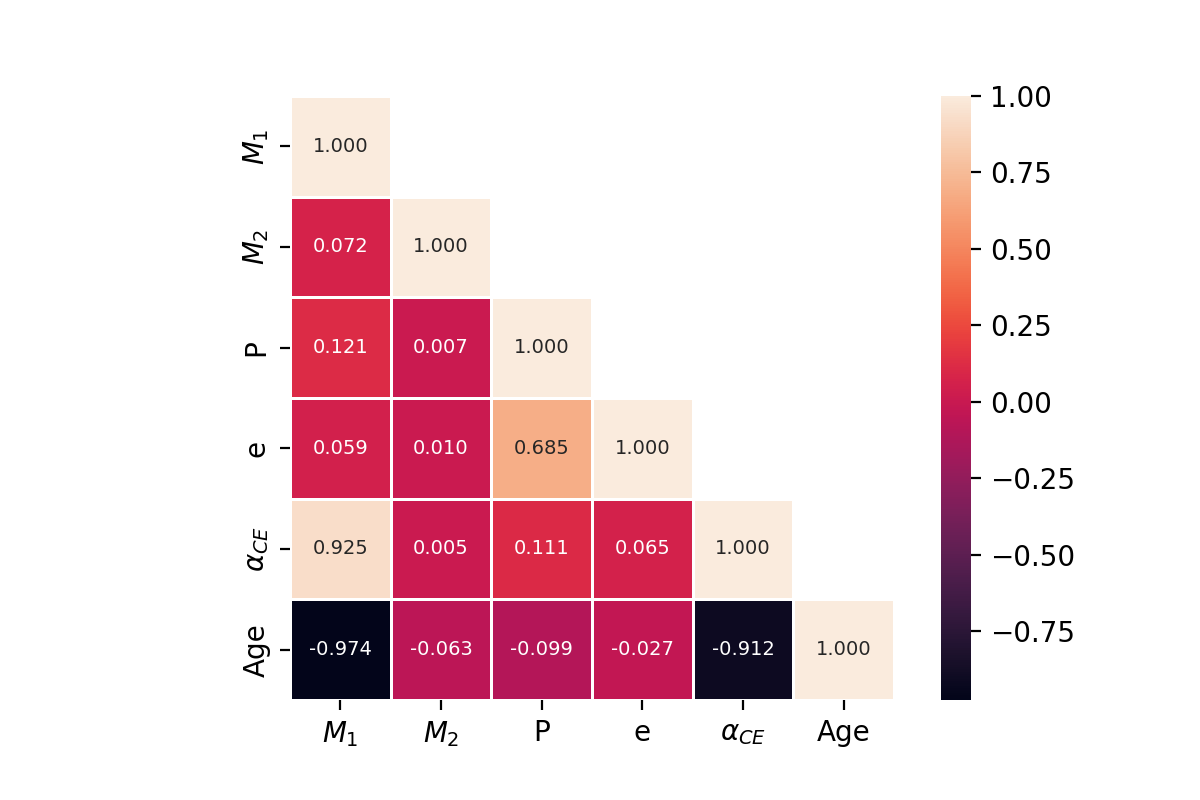}\par
\end{multicols}    
\begin{multicols}{2}    
    \includegraphics[width=0.95\columnwidth,trim=-90 -5 10 10, clip]{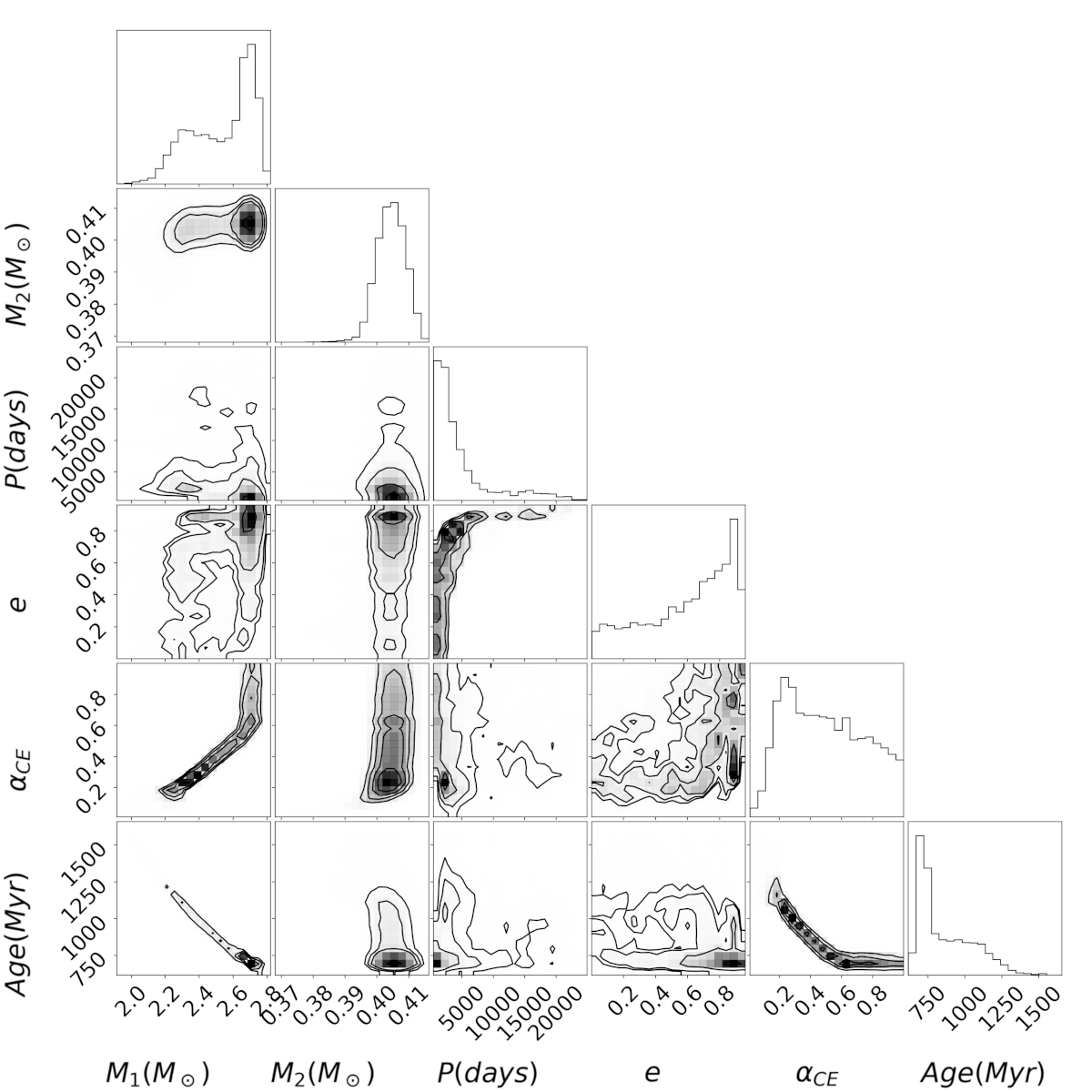}\par       
   \includegraphics[width=1.15\columnwidth,trim=80 15 -20 10, clip]{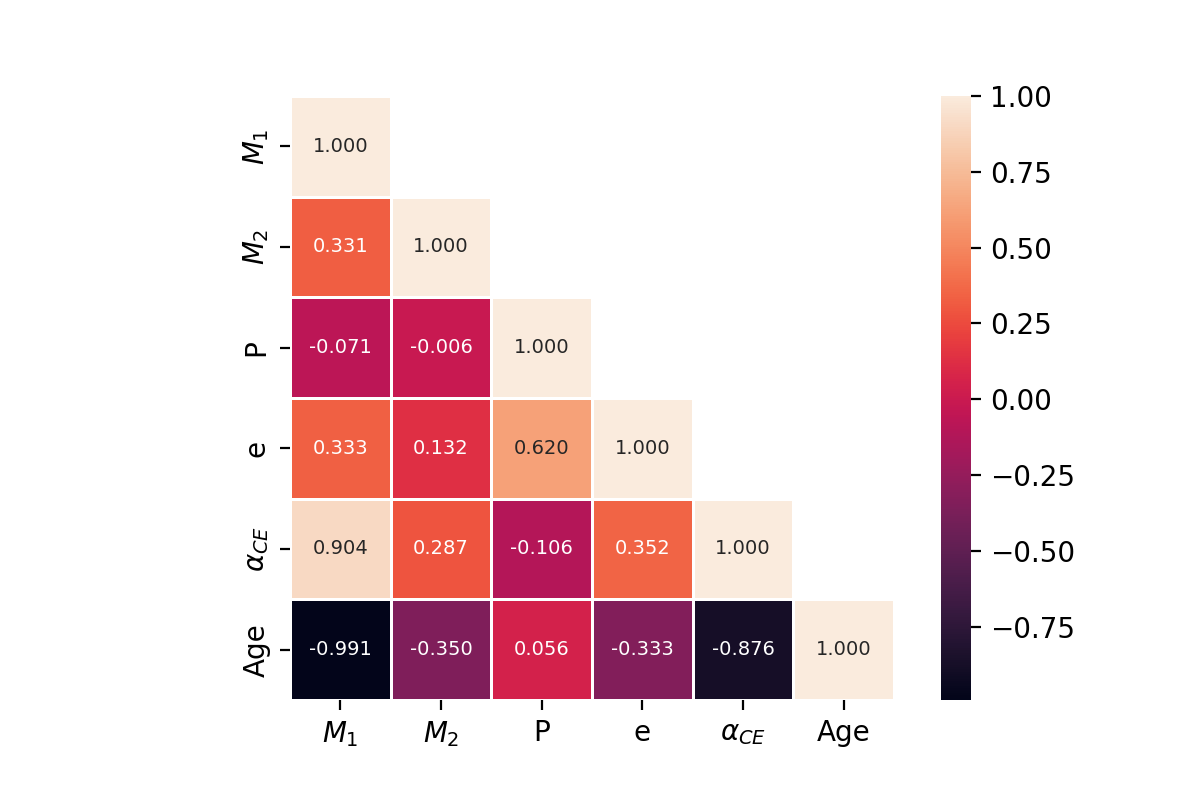}\par
\end{multicols}   
\begin{multicols}{2}   
    \includegraphics[width=0.95\columnwidth,trim=-90 -5 10 10, clip]{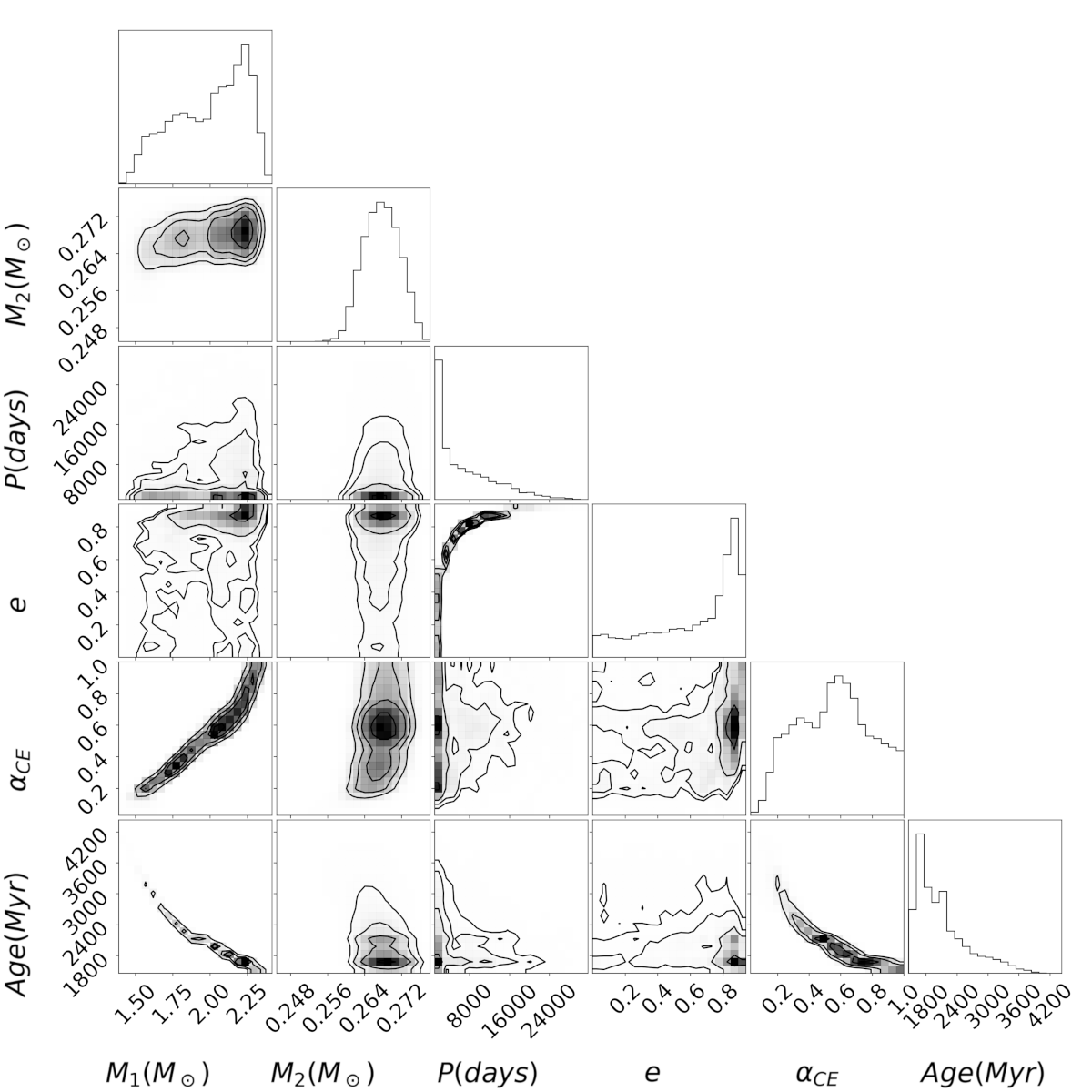}\par  
    \includegraphics[width=1.15\columnwidth,trim=80 15 -20 10, clip]{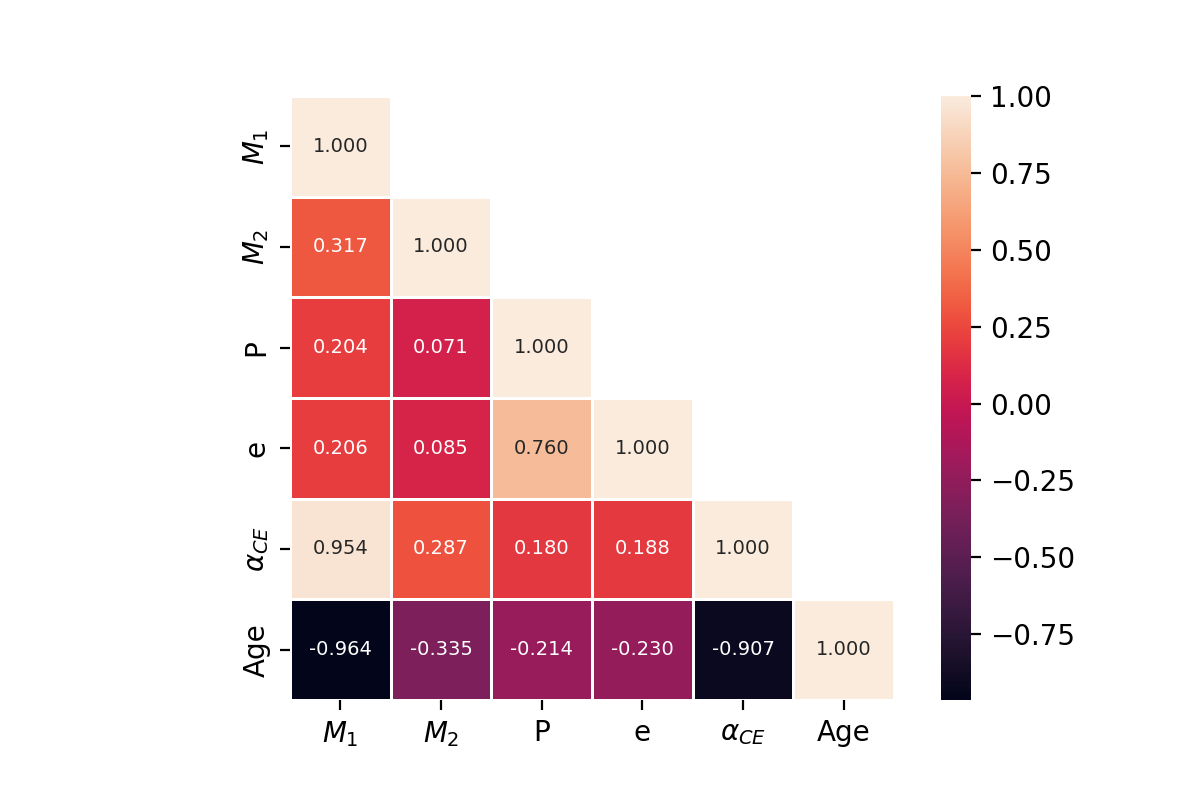}\par
\end{multicols}    
    \caption{Input space parameter obtained through our inverse population synthesis algorithm for three observed cases:  ZTF J041016.82-083419.5, ZTF J063808.71+091027.4, and ZTF J122009.98+082155.0 (top, middle, and bottom panels, respectively.}
    \label{f:3ztf}
\end{figure*}

 In Figure \ref{f:3ztf} we present the distribution of input parameters found for three representative cases and the associated corner plots (left panels), together with the corresponding correlation matrices (right panels): ZTF J041016.82-083419.5, ZTF J063808.71+091027.4, and ZTF J122009.98+082155.0. We remark that the distribution found for each parameter (histogram panels of each corner plot) is reasonable smooth, thus indicating that a stable solution has been found. 

A closer look reveals that, while the progenitor masses have a symmetric, near Gaussian distribution, that is not the case for the rest of parameters. In these cases, the asymmetric distribution implies that the average value (mean) does not agree with the most probable value (mode).  Inspection of the correlation matrices also revealed a strong correlation among some input parameters. That is the case of the common-envelope efficiency parameter $\alpha_{\rm CE}$ and the mass of the primary, $M_1$, where a positive correlation exits ($r\sim0.9\sim0.95$) for most of the systems. Similarly, a negative or anticorrelation arises between the age and $M_1$ and between the age and  $\alpha_{\rm CE}$ ($r\sim-0.9\sim-0.95$). These facts imply that the initial parameters are not independent variables. Therefore, if we were to adopt a solution set comprising the most probable or average value for each individual parameter, it would likely yield an incompatible solution with the observed values. Consequently, we have adopted the set of input values that correspond to the optimal solution, that is the solution that maximizes the joint probability distribution of the initial parameters (see Section \ref{ss:valalg}), for the remainder of the discussion.

In Table \ref{tab:min} we show the values corresponding to the optimal solution of our analyzed systems. Also, for sake of completeness, we include in the Appendix \ref{s:ann} Tables  \ref{tab:mean} and \ref{tab:moda}, corresponding to the mean and mode values, respectively. However, we should remark, as previously stated, that the initial conditions derived from the mean and mode values for each individual initial parameter, given the strong correlation among some of them, may not necessarily represent a solution of the binary system.

\begin{table*}[h!]
    \caption{Stellar input parameters for the optimal solution. The asterisks mark the only two systems for which their solution distance is above the observational error threshold. The last three columns represent, as percentage, the distance of the output parameters to the observed values, the distance once the shift in effective temperature is added, and the observational error threshold, respectively. See text for details.}
    \label{tab:min}
    \begin{center}
        \begin{tabular}{lccrccrccc}
            \hline
            Target & $M_{1,{\rm in}}$  & $M_{2,{\rm in}}$  & $P_{\rm in}$  & $e_{\rm in}$ & $\alpha_{\rm CE}$ & Age & $d_{\rm min}$ & $d'_{\rm min}$  & $\sigma_{\rm obs}$ \\
                             & (M$_{\odot}$) & (M$_{\odot}$) & (days) &  & & (Myr) & (\%) & (\%) & (\%) \\
            \hline
            \hline
            ZTF J041016.82-083419.5 & $1.316^{+0.032}_{-0.044}$ & $0.119^{+0.002}_{-0.002}$ & $11599^{+790}_{-11235}$ & $0.977^{+0.002}_{-0.253}$ & $0.977^{+0.023}_{-0.132}$ & $3870^{+448}_{-267}$ & 10.1 & 3.4 & 10.4 \\
            ZTF J051902.06+092526.4 & $1.878^{+0.030}_{-0.100}$ & $0.172^{+0.003}_{-0.005}$ & $3690^{+693}_{-3317}$ & $0.884^{+0.022}_{-0.720}$ & $0.651^{+0.138}_{-0.054}$ & $2226^{+266}_{-132}$ & 13.0 & 3.5 & 13.0 \\  
            ZTF J052848.24+215629.0 &  $3.189^{+0.047}_{-0.028}$ & $0.180^{+0.002}_{-0.001}$ & $10162^{+5916}_{-6282}$ & $0.820^{+0.051}_{-0.202}$ & $0.529^{+0.008}_{-0.019}$ & $1002^{+19}_{-18}$ & 9.8 & 2.1 & 11.0 \\ 
            ZTF J053708.26-245014.6 & $1.382^{+0.173}_{-0.053}$ & $0.197^{+0.004}_{-0.002}$ & $296^{+21622}_{-42}$ & $0.000^{+0.977}_{0.000}$ & $0.606^{+0.393}_{-0.028}$ & $4066^{+516}_{-1303}$ & 12.1 & 3.3 & 8.0 \\
            ZTF J061530.96+051041.8 & $2.198^{+0.071}_{-0.055}$ & $0.514^{+0.006}_{-0.007}$ & $1109^{+2045}_{-567}$ & $0.825^{+0.038}_{-0.177}$ & $0.687^{+0.076}_{-0.380}$ & $1287^{+85}_{-92}$ & 13.7 & 5.3 & 8.2 \\ 
            ZTF J063808.71+091027.4 & $2.707^{+0.026}_{-0.049}$ & $0.405^{+0.004}_{-0.005}$ & $2303^{+1038}_{-974}$ & $0.805^{+0.037}_{-0.195}$ & $0.898^{+0.102}_{-0.376}$ & $701^{+39}_{-15}$ & 9.6 & 1.2 & 9.2 \\ 
            ZTF J063954.70+191958.0 & $2.835^{+0.093}_{-0.069}$ & $0.206^{+0.005}_{-0.004}$ & $2281^{+28101}_{-729}$ & $0.739^{+0.207}_{-0.211}$ & $0.827^{+0.173}_{-0.291}$ & $705^{+26}_{-30}$  & 10.2 & 1.9 & 7.5 \\ 
            ZTF J064242.41+131427.6 & $2.431^{+0.042}_{-0.046}$ & $0.147^{+0.002}_{-0.002}$ & $1848^{+1037}_{-269}$ & $0.623^{+0.159}_{-0.104}$ & $0.642^{+0.356}_{-0.049}$ & $960^{+40}_{-29}$ & 7.5 & 3.2 & 7.8 \\ 
            ZTF J065103.70+145246.2 & $1.612^{+0.128}_{-0.197}$ & $0.237^{+0.004}_{-0.005}$ & $1555^{+10215}_{-686}$ & $0.597^{+0.313}_{-0.510}$ & $0.205^{+0.064}_{-0.047}$ & $2503^{+1118}_{-446}$ & 11.1 & 2.4 & 10.4 \\ 
            ZTF J070458.08-020103.3 &  $1.344^{+0.020}_{-0.027}$ & $0.338^{+0.004}_{-0.004}$ & $777^{+130}_{-101}$ & $0.042^{+0.185}_{-0.040}$ & $0.087^{+0.012}_{-0.012}$ & $4635^{+293}_{-175}$ & 11.5 & 3.5 & 7.4 \\ 
            ZTF J071759.04+113630.2 & $1.941^{+0.045}_{-0.044}$ & $0.289^{+0.005}_{-0.006}$ & $923^{+605}_{-214}$ & $0.363^{+0.329}_{-0.164}$ & $0.449^{+0.325}_{-0.061}$ & $1552^{+0}_{-20}$  & 13.4 & 4.0 & 9.3 \\ 
            ZTF J071843.68-085232.1 & $3.603^{+0.118}_{-0.090}$ & $0.297^{+0.007}_{-0.005}$ & $20746^{+30474}_{-10428}$ & $0.872^{+0.058}_{-0.090}$ & $0.317^{+0.011}_{-0.014}$ & $459^{+18}_{-17}$ & 13.3 & 3.9 & 9.4 \\ 
            ZTF J080542.98-143036.3 & $1.320^{+0.494}_{-0.055}$ & $0.288^{+0.003}_{-0.005}$ & $5050^{+403}_{-2624}$ & $0.944^{+0.005}_{-0.034}$ & $0.388^{+0.429}_{-0.028}$ & $4004^{+606}_{-2540}$ & 11.6 & 2.4 & 9.9 \\ 
            ZTF J094826.35+253810.6 & $1.094^{+0.123}_{-0.036}$ & $0.165^{+0.001}_{-0.002}$ & $627^{+756}_{-82}$ & $0.466^{+0.230}_{-0.185}$ & $0.141^{+0.045}_{-0.032}$ & $7017^{+783}_{-1959}$ & 10.8 & 4.3 & 12.4 \\ 
            ZTF J102254.00-080327.3 &  $2.659^{+0.087}_{-0.085}$ & $0.396^{+0.009}_{-0.006}$ & $864^{+403}_{-327}$ & $0.540^{+0.121}_{-0.534}$ & $0.882^{+0.069}_{-0.033}$ & $1733^{+67}_{-61}$ & 14.2 & 4.7 & 10.1 \\ 
            ZTF J102653.47-101330.3 & $1.294^{+0.090}_{-0.039}$ & $0.103^{+0.001}_{-0.001}$ & $687^{+1606}_{-198}$ & $0.795^{+0.108}_{-0.106}$ & $0.942^{+0.058}_{-0.176}$ & $3469^{+370}_{-639}$ & 2.6 & 9.4 & 10.1 \\ 
            ZTF J104906.96-175530.7 & $1.645^{+0.089}_{-0.052}$ & $0.194^{+0.004}_{-0.004}$ & $254^{+32}_{-30}$ & $0.003^{+0.309}_{-0.003}$ & $0.852^{+0.120}_{-0.117}$ & $2261^{+195}_{-251}$  & 10.1 & 1.8 & 8.4 \\ 
            ZTF J122009.98+082155.0 & $2.238^{+0.045}_{-0.073}$ & $0.268^{+0.004}_{-0.005}$ & $1686^{+247}_{-162}$ & $0.067^{+0.455}_{-0.066}$ & $0.645^{+0.317}_{-0.075}$ & $1641^{+105}_{-32}$ & 12.5 & 3.3 & 9.0 \\ 
            ZTF J125620.57+211725.$8^{\ast}$ & $1.459^{+0.045}_{-0.040}$ & $0.102^{+0.000}_{+0.000}$ & $974^{+28}_{-76}$ & $0.072^{+0.093}_{-0.070}$ & $0.838^{+0.098}_{-0.049}$ & $5033^{+217}_{-239}$ & 1.6 & 11.7 & 7.0 \\ 
            ZTF J130228.34-003200.2 & $3.438^{+0.220}_{-0.105}$ & $0.174^{+0.004}_{-0.004}$ & $29577^{+11907}_{-2758}$ & $0.915^{+0.007}_{-0.025}$ & $0.466^{+0.036}_{-0.043}$ & $1041^{+41}_{-44}$ & 10.8 & 3.7 & 9.0 \\ 
            ZTF J134151.70-062613.9 &  $1.793^{+0.018}_{-0.018}$ & $0.122^{+0.001}_{-0.002}$ & $1180^{+57}_{-74}$ & $0.697^{+0.017}_{-0.027}$ & $0.756^{+0.064}_{-0.069}$ & $1632^{+42}_{-39}$ & 11.0 & 5.9 & 21.6 \\ 
            ZTF J140036.65+081447.4 & $2.303^{+0.163}_{-0.121}$ & $0.225^{+0.006}_{-0.006}$ & $15143^{+4218}_{-13521}$ & $0.957^{+0.005}_{-0.163}$ & $0.933^{+0.067}_{-0.528}$ & $1272^{+192}_{-148}$ & 13.0 & 3.7 & 8.4 \\
            ZTF J140423.86+065557.7 & $2.918^{+0.059}_{-0.055}$ & $0.399^{+0.006}_{-0.006}$ & $9034^{+635}_{-732}$ & $0.886^{+0.006}_{-0.014}$ & $0.491^{+0.024}_{-0.054}$ & $838^{+41}_{-32}$ & 16.3 & 6.9 & 7.9 \\ 
            ZTF J140537.34+103919.$0^{\ast}$ & $0.908^{+0.026}_{-0.036}$ & $0.096^{+0.002}_{-0.001}$ & $2702^{+227}_{-2033}$ & $0.928^{+0.005}_{-0.124}$ & $0.696^{+0.210}_{-0.075}$ & $8814^{+1397}_{-838}$ & 21.0 & 28.7 & 8.5 \\ 
            ZTF J140702.57+211559.7 &  $1.788^{+0.058}_{-0.087}$ & $0.257^{+0.005}_{-0.005}$ & $33240^{+3390}_{-32227}$ & $0.981^{+0.002}_{-0.196}$ & $0.585^{+0.058}_{-0.087}$ & $2317^{-268}_{-121}$ & 11.1 & 2.1 & 10.6 \\ 
            ZTF J162644.18-101854.3 & $1.144^{+0.016}_{-0.030}$ & $0.206^{+0.004}_{-0.004}$ & $1432^{+152}_{-248}$ & $0.578^{+0.038}_{-0.021}$ & $0.100^{+0.021}_{-0.009}$ & $6497^{+617}_{-306}$  & 10.1 & 3.6 & 10.9 \\ 
            ZTF J163421.00-271321.7 & $1.731^{+0.077}_{-0.128}$ & $0.126^{+0.001}_{-0.002}$ & $780^{+90}_{-77}$ & $0.208^{+0.115}_{-0.206}$ & $0.426^{+0.079}_{-0.058}$ & $2524^{+826}_{-185}$ & 23.9 & 11.8 & 18.0 \\ 
            ZTF J164441.18+243428.2 & $1.515^{+0.039}_{-0.087}$ & $0.096^{+0.002}_{-0.001}$ & $339^{+202}_{-35}$ & $0.345^{+0.233}_{-0.089}$ & $0.932^{+0.067}_{-0.133}$ & $3020^{+540}_{-223}$ & 9.0 & 9.2 & 12.0 \\ 
            ZTF J180256.45-005458.3 & $1.591^{+0.017}_{-0.025}$ & $0.147^{+0.001}_{-0.001}$ & $484^{+684}_{-230}$ & $0.797^{+0.051}_{-0.028}$ & $0.941^{+0.057}_{-0.051}$ & $2189^{+75}_{-68}$ & 7.8 & 2.9 & 10.7 \\ 
            ZTF J195456.71+101937.5 & $1.371^{+0.055}_{-0.007}$ & $0.443^{+0.004}_{-0.004}$ & $2011^{+598}_{-567}$ & $0.698^{+0.044}_{-0.113}$ & $0.122^{+0.022}_{-0.022}$ & $4249^{+71}_{-509}$ & 6.6 & 2.6 & 9.2 \\
            \hline
        \end{tabular}
    \end{center}
\end{table*}

For the set of optimal solutions we computed the distance to the observed values following equation \ref{eq:distance}. In Table \ref{tab:min} these distances are shown as a percentage under column $d_{\rm min}$. Likewise, we display in the last column of the table, the observational error $\sigma_{\rm obs}$; calculated as the sum in quadrature from the error estimates presented in Table \ref{tab:ztf} for $M_1$, $M_2$, $T_{\rm eff,1}$ and  $T_{\rm eff,2}$, and, as a result of error propagation,  2.5\% for the period. This  $\sigma_{\rm obs}$ is adopted as our threshold error in the computation of solutions. An initial analysis revealed that for only 11 systems, the set of solutions provide a distance below the observed error threshold. 

\begin{figure*}
    \includegraphics[width=0.98\textwidth,trim=0 0 0 0, clip]{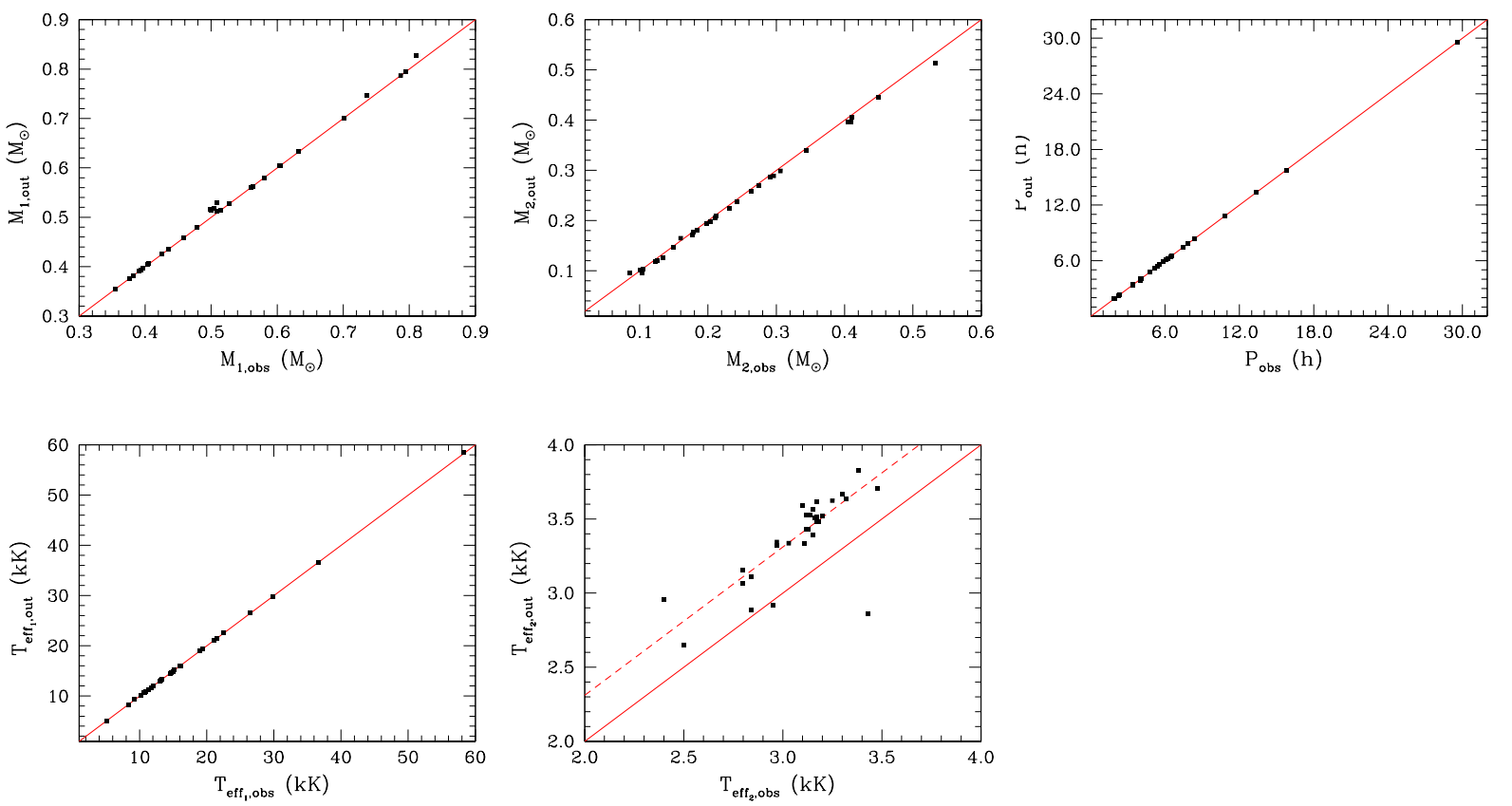}
    \caption{Parameter comparison of output values for the solution that provides the minimum distance, $(M_1, M_2, T_{\rm eff,1}, T_{\rm eff,2}, P)_{\rm out}$, with respect to the  observed values, $(M_1, M_2, T_{\rm eff,1}, T_{\rm eff,2}, P)_{\rm obs}$. As it can be seen, all parameters perfectly agree except $T_{\rm eff,2}$, which shows a clear shift between the output versus observed values.  See text for details.}
    \label{f:outobs}
\end{figure*}

To investigate this result in more detail, we have depicted in Figure \ref{f:outobs} the output  set of values for our optimal solution, $(M_1, M_2, T_{\rm eff,1}, T_{\rm eff,2}, P)_{\rm out}$, with respect to the corresponding  observed values, $(M_1, M_2, T_{\rm eff,1}, T_{\rm eff,2}, P)_{\rm obs}$. As a visual aid, we marked (solid black line) the 1:1 relationship between parameters. A practically perfect agreement is achieved for all parameters except for $T_{\rm eff,2}$. In that case, the obtained solution is systematically hotter than the observed value. This issue was reported in \citet[][see Fig. 5.]{Brown2023}. 

From this analysis we conclude than the main source of error arises from the discrepancy in the effective temperature of the secondary. We can account for this problem by adopting an {\sl ad hoc} shift of 310\,K (dashed black line) in the observed effective temperature estimates or, equivalently, in theoretical model. The new distances for equation \ref{eq:distance} we would obtain in this case are written in Table \ref{tab:min} under the column $d'_{\rm min}$. With the exception of two cases, all the others now exhibit a distance below the observed threshold error. These two (ZTF J125620.57+211725.8 and ZTF J140537.34+103919.0) and  have been reported in \citet{Brown2023} as systems of special interest. 

ZTF J125620.57+211725.8 was originally identified in \cite{Rebassa2021} as an unresolved binary system formed by a extremely low mass (below $0.2\,$M$_\odot$) white dwarf. On the other hand, the mass estimate of the white dwarf by \cite{Brown2023} was notably larger, specifically $0.48\pm0.01\,$M$_\odot$. Our inverse population synthesis study reveals that is not possible to find a set of progenitor parameters that reproduce the observed fitted values presented in \cite{Brown2023}. Thus, we reinforce the idea of conducting spectroscopic follow-up alongside a detailed analysis of binary evolution for this particular system.

Regarding ZTF J140537.34+103919.0, this system was reported in \cite{Brown2023}
as a binary with a  possible sub-stellar companion, whose secondary star exhibits an effective temperature hotter that expected for its mass. The same authors suggested the possibility of it being a triple system to explain their observed peculiarities. Nevertheless, a thorough analysis of this system is deserved to determine the progenitors and understand their past history.

Finally, it is important to note that, although the data quality of the sample analyzed here is sufficient to yield observational uncertainties on the order of a few percent, this level of uncertainty slightly surpasses the ideal conditions (see Section \ref{ss:valalg}) required to accurately recover the input parameters. Consequently, some caution should be exercised in the interpretation of the results presented here and in the forthcoming sections.

\subsection{Global properties of the progenitors}
\label{ss:global}

\begin{figure*}
\centering
    \includegraphics[width=0.8\textwidth,trim=0 0 0 0, clip]{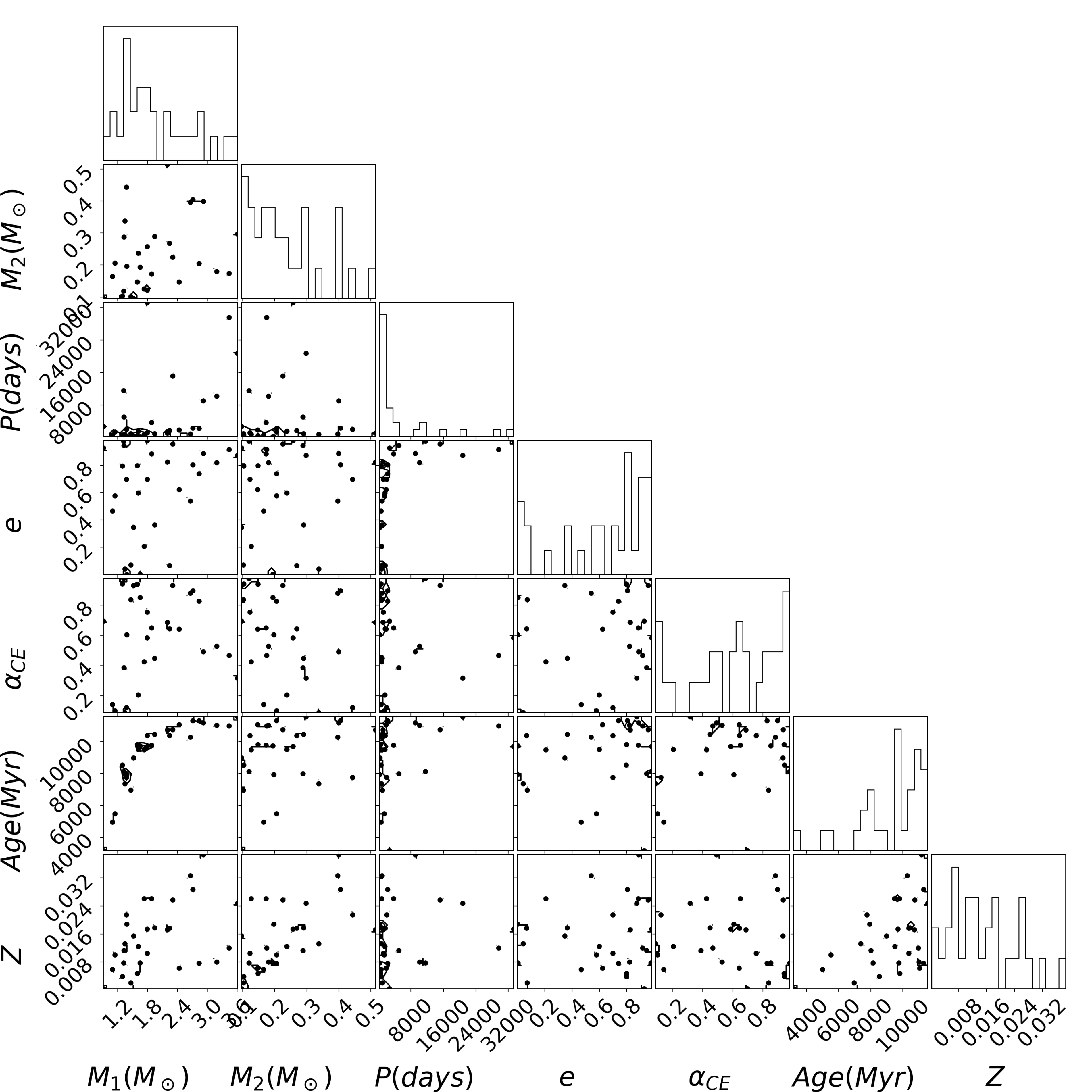}
    \includegraphics[width=0.75\textwidth,trim=0 0 0 0, clip]{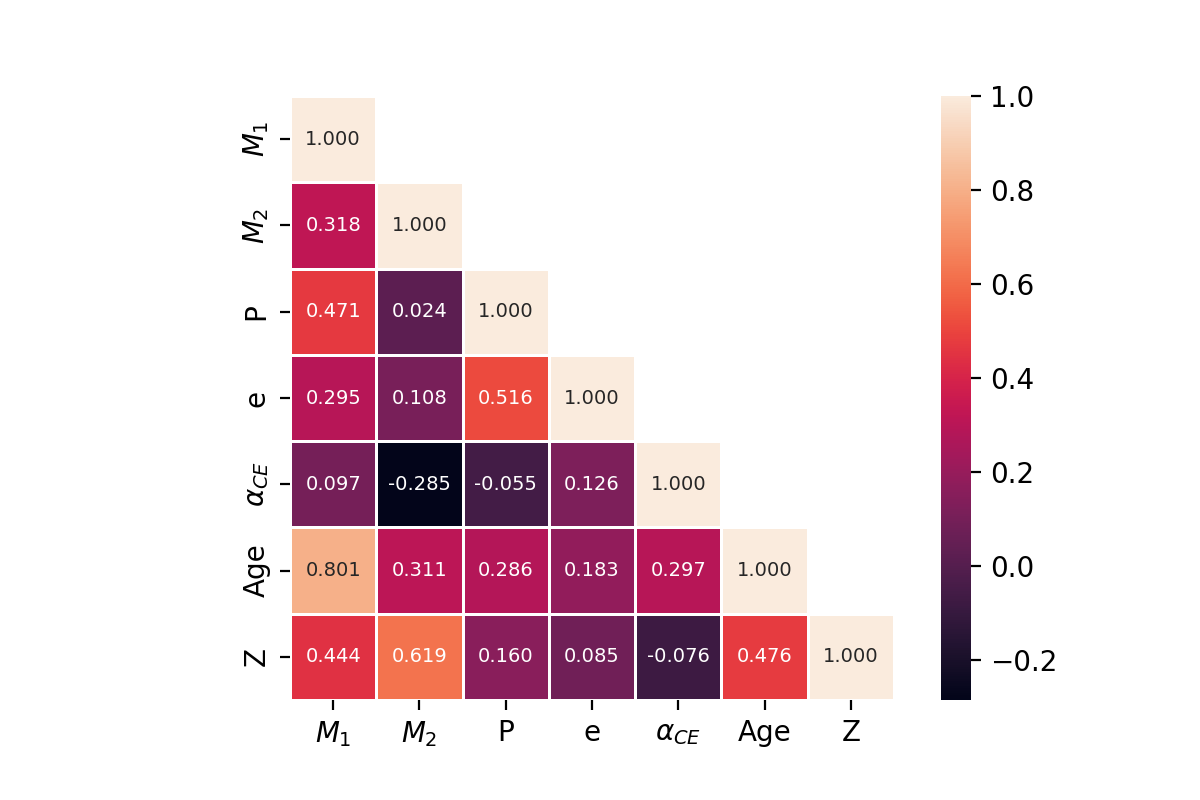}
    \caption{Distribution of parameters for the entire sample.}
    \label{f:entire}
\end{figure*}

Once each individual system was analyzed, we studied the properties of the entire sample (28 systems which have compatible solutions with the observed errors; see Table \ref{tab:min}). Recall that we adopted as progenitor solutions the set of parameters that resulted in the maximum joint probability within the observed parameter space. In Figure \ref{f:entire} we display a corner-plot for the progenitor parameters (left panel) and the respective correlation matrices (right panel). As input parameters we have used the masses of the progenitor stars, $M_1$ and $M_2$, the eccentricity and orbital period, $e$ and $P$, the common envelope efficiency parameter, $\alpha_{\rm CE}$, the age of the system, $T_{\rm age}$, and we have also added the metallicity, $Z$ as computed in Section \ref{s:obsa}.

Some correlations between the parameters are detected. First, a clear anticorrelation, $r=-0.801$, arises between the total age and the mass $M_1$. As expected from stellar evolution theory, primary low-mass  progenitors (roughly around 1\,\Msun),  spend more time until they overfill their Roche-lobes. This natural effect is also increased if we consider that we have estimated low-metallicities for those low-mass primaries. 

\begin{figure}
	\includegraphics[width=0.95\columnwidth,trim=10 70 40 60, clip]{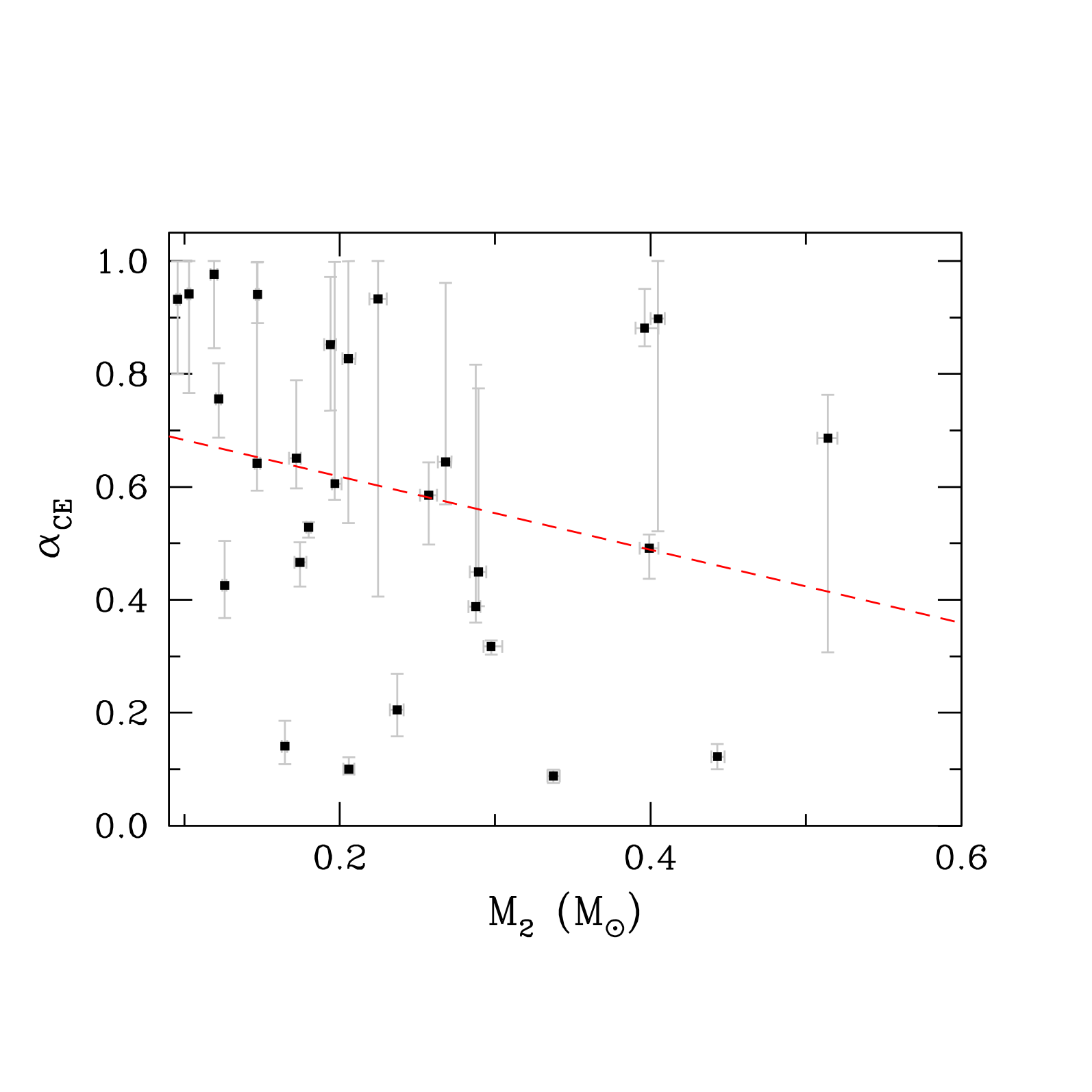}
    \caption{Common envelope efficiency parameter, $\alpha_{\rm CE}$, as a function of the mass of the secondary. The decreasing linear trend is marked by the dashed red line.}
    \label{f:m2alpha}
\end{figure}

Second, a mild anticorrelation, $r=-0.285$, is shown between the  $\alpha_{\rm CE}$ parameter and the mass of the secondary. For a better analysis, we plot in Figure \ref{f:m2alpha}  $\alpha_{\rm CE}$ versus $M_2$ with its corresponding error bars and marked as a dashed red line the linear decreasing trend. The confidence interval for the slope is $\left[-1.656, 0.359\right]\,$(\Msun)$^{-1}$ at a 95\% confidence level, indicating that a null or even positive slope cannot be entirely excluded. However, the relation found suggests that the largest values of the common-envelope efficiency parameter correspond to low-mass secondaries, although not all of these systems present a large $\alpha_{\rm CE}$. On the other hand, more massive secondaries present low-values of $\alpha_{\rm CE}$. Moreover, the analysis of the distribution of  the common-envelope efficiency parameter reveals that there is not a unique value (on the contrary there is a wide range from $\alpha_{\rm CE}=0.087$ to 0.977) capable of reproducing the entire sample. The average value of the common-envelope  efficiency for the whole sample result to be $\langle\alpha_{\rm CE}\rangle=0.588$, which is slightly larger than in previous analysis of post-common envelope WDMS binaries that assigned a low efficiency parameter  $\alpha_{\rm CE}\sim0.2-0.4$ \citep[e.g.][]{Zorotovic2010,Toonen2013,Camacho2014}. This is partially due to the fact that some of the systems with the lowest secondary-mass require a higher efficiency, in some cases  $\alpha_{\rm CE}$ is nearly as high as 1. For instance, this is the case for ZTF J164441.18+243428.2 and ZTF J102653.47-101330.3, both with $M_2<0.11\,$\Msun\ and with $\alpha_{\rm CE}= 0.932$ and 0.942, respectively. This result also suggests that for low-mass secondaries, the inclusion of other sources of energy, in addition to the orbital energy, such as the thermal envelope energy or internal energy due to dissociation and ionization energy of the envelope, may be considered \citep{Rebassa2012}. 

The aforementioned suggestion seems to be in line with previous analysis by \cite{DeMarco2011,Davis2011}. These authors hypothesized that lower mass companions require more time than a stellar dynamical time to spiral into the core of the giant. This prolonged process  enables the giant to use its internal thermal energy to facilitate the unbinding of its envelope. This effect should be more evident in the case of brown dwarf companions instead of M~dwarfs. However, in a recent analysis of a sample of close detached white dwarf plus brown dwarf binaries \citet{Zorotovic2022} indicated that these systems can be reconstructed with a low-efficiency parameter without the need to include  other sources of internal energy for expelling the envelope. However, it should be noted that this result is achieved by imposing the condition that the age of the analyzed systems should be above a minimum age derived from the literature. If this condition is omitted, larger values of the efficiency parameter are recovered, indicating in that case the contribution of a certain amount of internal energy.

It is worth mentioning here that our algorithm does not impose any restrictions on the input parameters, which include the total age of the system. Furthermore, since our observational sample includes a precise estimation of the temperatures of both stars, the algorithm is thus able to properly constrain the total age of the system.

Continuing our analysis of the output parameters, we also observed a certain anticorrelation between the common envelope efficiency and the age of the system. Large values of $\alpha_{\rm CE}$ are linked to young ages. However, the opposite is not entirely true, as low efficiency values are associated with both young and old systems. This fact implies that, at least for the sample analyzed here, there is no clear anticorrelation between the common envelope efficiency and the mass ratio, $q=M_2/M_1$, as proposed by \citet{DeMarco2011}.

\begin{figure}
	\includegraphics[width=0.95\columnwidth,trim=10 70 40 60, clip]{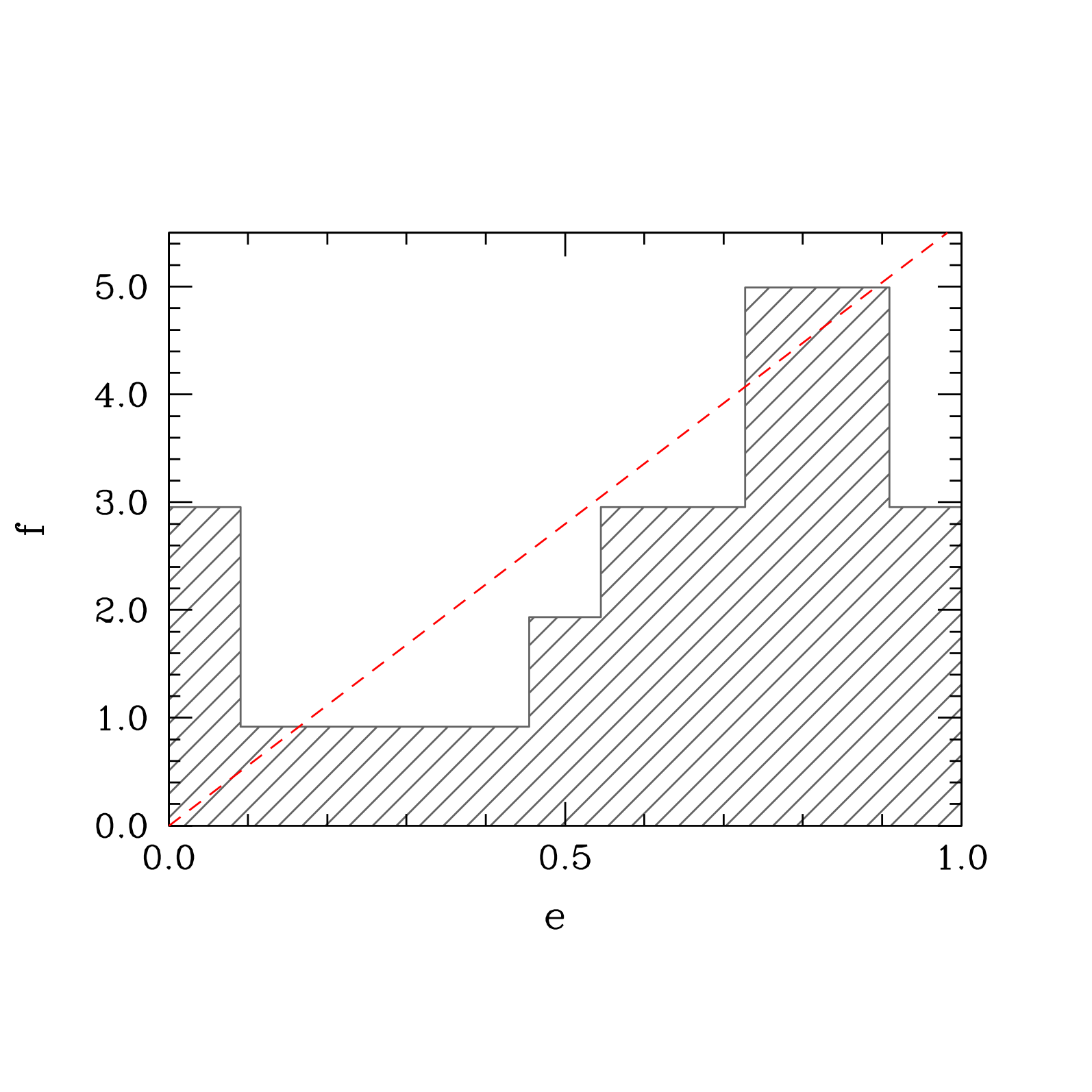}
    \caption{Initial eccentricity distribution for the systems analyzed in this work. Despite the scarcity of the data, the distribution seems to be consistent with the canonical thermal distribution (red line), $f=2e$, although some initially already circularized systems appear in the sample.}
    \label{f:eccentricity}
\end{figure}

Another noteworthy result of the set of solutions found by our algorithm concerns to the eccentricity distribution. In Figure \ref{f:eccentricity} we plot the initial eccentricity distribution for the systems analyzed here. For visual analysis we depict the canonical thermal distribution (red dashed line), $f=2e$. We should be aware that a larger phase space at higher eccentricities is available for the random walkers (for instance, binary systems circularized at periastron upon Roche lobe overflow, allowing for a much wider range of orbital separations). However, given that the optimal solution corresponding to the product of the marginal distributions is maximized, individual biases can, to some extent, be compensated. For instance, binaries with shorter initial periods are naturally more prone to Roche lobe overflow and thus tend to interact earlier (as shown in the marginal distributions of the corner plots in Figure \ref{f:val}). These effects therefore compensate for the larger phase space available at higher eccentricities when determining the optimal solution for the system. We conclude that, as long as the observational errors are small, the optimal solution can be considered a reasonable estimate, and therefore the distribution obtained from the analyzed sample reflects the properties of that sample. Hence, despite the fact that our sample cannot be considered complete by far, and the existence of some initially already circularized systems ($\sim10\%$), the eccentricity distribution found in this work sample seems to be compatible with the thermal one.  A Kolmogorov-Smirnov test yields a $p$-value of 0.2054, failing to reject the null hypothesis that the data follows the thermal distribution. This result suggests that close binary systems exhibit an initially thermalized distribution of eccentricities, rather than a uniform or even superthermal distribution found in wide binaries \citep[e.g.,][]{Geller2019,Hwang2022}.

\begin{figure}
	\includegraphics[width=0.95\columnwidth,trim=10 70 40 60, clip]{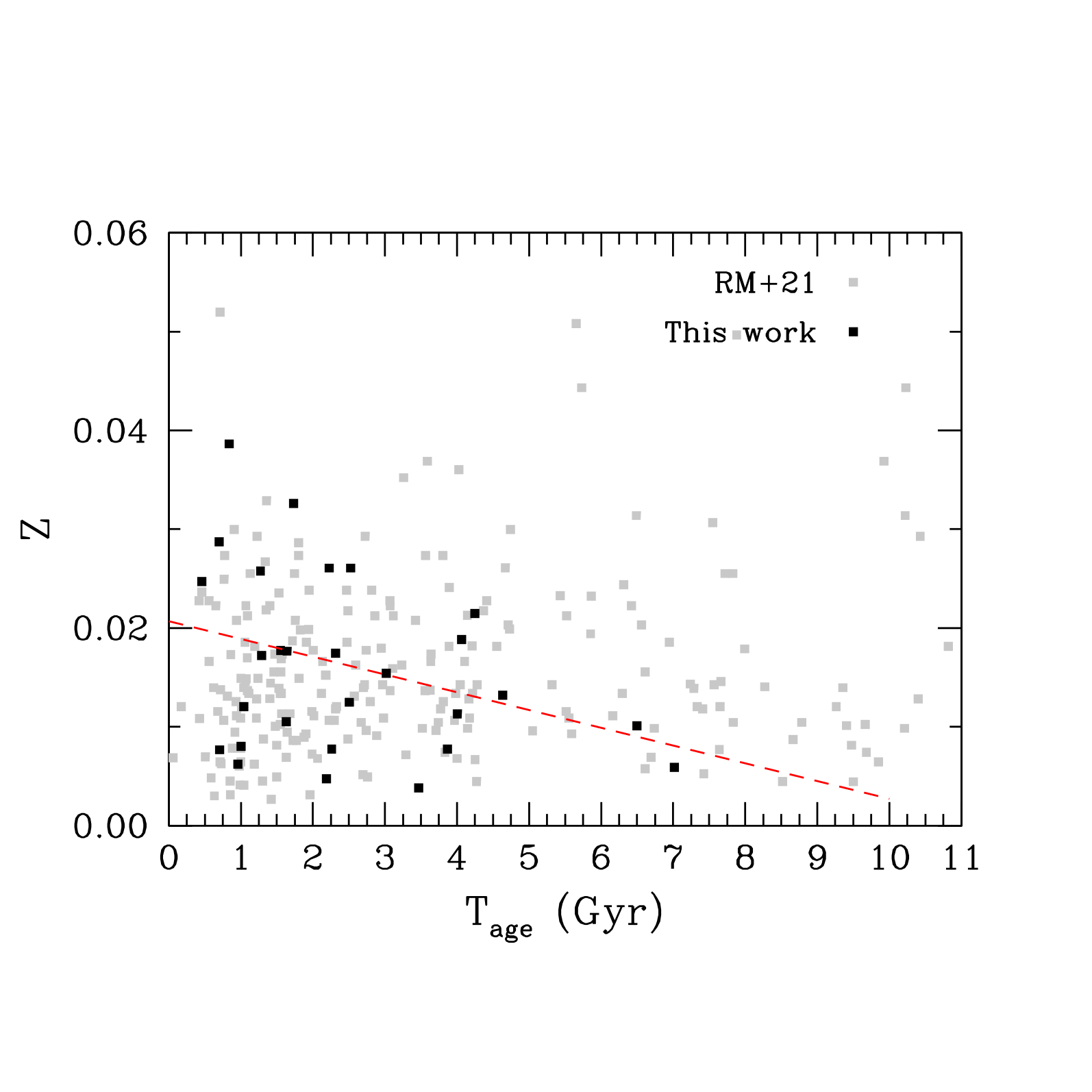}
    \caption{Age-metallicity relationship for the objects analyzed in this work (black points). For visual comparison we also show the linear trend of our sample (red dashed line) and the sample of wide WDMS binaries from \citep[][gray dots, RM+21]{Rebassa2021}. }
    \label{f:agemetal}
\end{figure}

Finally, our study of the sample of eclipsing binary systems from \citet{Brown2023} has allowed us to estimate the metallicities (see Section \ref{s:obsa}) and, through our reconstruction algorithm, the ages of such binary systems. Hence, we can analyze the resulting age-metallicity distribution for objects in our sample. In Figure \ref{f:agemetal} we show the distribution obtained (black dots) along with a best-fit linear trend (red dashed line). For visual comparison we also plot the age-metallicity values (gray dots) for the sample of wide WDMS binaries from \citet{Rebassa2021}. The dispersion in metallicities presented in our sample for younger ages (up to $\sim3\,$Gyr) is in agreement with that presented in \citet{Rebassa2021}. However, a noticeable trend towards lower metallicities is shown in our sample. Although the 95\% confidence interval for the slope, $\left[-3\,870, 799\right]\,$(yr$^{-1}$), reveals that it is not entirely incompatible with a null slope. In any case, the lack of older high-metallicity objects in our sample, in contrast to the sample of \citet{Rebassa2021}, where this dispersion in metallicities is consistently maintained across all ages, can be attributed to the shorter main-sequence lifetime of these objects compared to their low-metallicity counterparts.  Hence, they can enter the common-envelope phase earlier and spend more time cooling, becoming too faint to be detected, or even lead to a merger of the binary system. A complete volume-limited sample would provide valuable insight into this issue.

\section{Conclusions}

We have developed a binary stellar reconstruction algorithm based on inverse population synthesis techniques. The algorithm, employing adaptive random walks, efficiently explores the input parameter space and identifies a set of solutions. This set consists of initial parameters that, when the binary system evolves, result in output parameters compatible with the observed ones within a certain error threshold.

The algorithm was validated with the aid of \texttt{MRBIN}, a population synthesis code devoted to binary evolution and based on \texttt{BSE} code. We verified that our algorithm can identify a stable solution after computing approximately $10^6$ points per system and around $100$ random walks in parallel, representing a reduced computational time. At the same time, this solution is obtained without the need to impose any restriction on  the input variables.

Once validated, we applied our algorithm to a set of eclipsing WDMS binaries detected by the ZTF survey. The high-precision parameters of this set of observed systems allows us to obtain an accurate estimate of the progenitor parameters without the need to impose any external constraints. Out of the 30 systems analyzed, we found a valid solution, meaning that the distance between the output parameters and the observed values is below the error threshold, for 28 of them.

The input parameter space for those 28 systems is defined by the masses of the progenitor stars, $M_1$ and $M_2$, the eccentricity and orbital period, $e$ and $P$, the common envelope efficiency parameter, $\alpha_{\rm CE}$, and the age of the system, $T_{\rm age}$. We have also made a priori estimates of the metallicity, $Z$, computed through interpolation of \texttt{PARSEC} sequences, and kept fixed throughout the application of the algorithm. The output parameters correspond to the masses and effective temperatures of the observed objects, as well as their orbital periods: $(M_1, M_2, T_{\rm eff,1}, T_{\rm eff,2}, P){\rm obs}$. The output parameter values provided by our algorithm perfectly match the observed ones, except for $T{\rm eff,2}$. In this case, a shift of 310\,K is adopted due to a calibration issue in the synthetic evolutionary tracks.

The analysis of the global properties has revealed significant trends and correlations among certain parameters. They can be summarized in the following points:

\begin{itemize}
\item A high degeneracy exists in the input parameter space, although the excellent precision of the observational data to which we apply our algorithm allows us to constrain the input values to a reasonable extent.
\item Not a universal value of the common envelope efficiency, $\alpha_{\rm CE}$ allows for the reconstruction of all systems; instead, a wide range of values is needed. In fact, the average $\alpha_{\rm CE}$ value we find is $\simeq$0.6, which is slightly higher than the 0.3--0.4 values generally assumed in previous works.
\item The $\alpha_{\rm CE}$ parameter exhibits a mild anticorrelation with the secondary mass. Although no internal energy is required for the systems analyzed here, larger values found for the lowest secondary masses suggest that internal energy may be considered in this regime. Furthermore, the extrapolation of this result also suggests that the inclusion of this extra source of energy can be present and may even play a major role in systems with brown-dwarf companions.
\item Initial eccentricities are compatible with a thermal distribution, although a small fraction ($\sim10$\%) of systems are initially circularized. 
\item The age-metallicity relation found in this work resembles that obtained for wide binary systems, although in our sample of post-common-envelope binaries, there is a lack of old and high-metallicity objects, probably due to an observational bias against these objects.
\end{itemize}

Although preliminary, the results presented in this work highlight important trends. A more detailed and robust analysis, incorporating a complete sample along with the effects of selection biases and prior assumptions, will be necessary to fully confirm and refine these findings.

\begin{acknowledgements}
We acknowledge the positive and valuable comments from our anonymous referee. We acknowledge support from MINECO under the PID2020-117252GB-I00 and PID2023-148661NB-I00 grant and by the AGAUR/Generalitat de Catalunya grant SGR-386/2021. 
\end{acknowledgements}

\bibliographystyle{aa}
\bibliography{IPST}
\section{Annex}
\label{s:ann}

Tables \ref{tab:mean} and \ref{tab:moda} correspond to the mean (average) and mode (most probable value) solution for the stellar input parameters of the sample analyzed in this work. It should be recalled that, due to the correlation between the parameters, the set of input values for a certain system does not guarantee that it should be a solution. That is, after evolving the system with these initial set of parameters, the output values can be in disagreement with the observed values (see Section \ref{s:resu}).

\begin{table*}[h!]
    \caption{Mean (average) solution for the stellar input parameters.}
    \label{tab:mean}
    \begin{center}
        \begin{tabular}{lccrccr}
            \hline
            Target & $M_{1,{\rm in}}$  & $M_{2,{\rm in}}$  & $P_{\rm in}$  & $e_{\rm in}$ & $\alpha_{\rm CE}$ & Age  \\
                             & (M$_{\odot}$) & (M$_{\odot}$) & (days) &  & & (Myr) \\
            \hline
            \hline
ZTF	J041016.82-083419.5	&	1.235	&	0.119	&	3916.2	&	0.891	&	0.789	&	5078.4	\\
ZTF	J051902.06+092526.4	&	1.544	&	0.171	&	2738.3	&	0.716	&	0.503	&	4174.6	\\
ZTF	J052848.24+215629.0	&	3.129	&	0.180	&	10918.2	&	0.668	&	0.401	&	1020.0	\\
ZTF	J053708.26-245014.6	&	1.397	&	0.198	&	12082.5	&	0.871	&	0.731	&	4293.7	\\
ZTF	J061530.96+051041.8	&	2.171	&	0.514	&	3980.3	&	0.723	&	0.463	&	1344.5	\\
ZTF	J063808.71+091027.4	&	2.533	&	0.404	&	4780.8	&	0.601	&	0.500	&	867.2	\\
ZTF	J063954.70+191958.0	&	2.733	&	0.206	&	11332.5	&	0.691	&	0.534	&	775.0	\\
ZTF	J064242.41+131427.6	&	2.306	&	0.147	&	4438.9	&	0.603	&	0.573	&	1110.1	\\
ZTF	J065103.70+145246.2	&	1.538	&	0.237	&	5187.4	&	0.705	&	0.214	&	3065.4	\\
ZTF	J070458.08-020103.3	&	1.273	&	0.337	&	6390.6	&	0.650	&	0.080	&	5708.9	\\
ZTF	J071759.04+113630.2	&	1.755	&	0.289	&	5901.1	&	0.698	&	0.493	&	2182.7	\\
ZTF	J071843.68-085232.1	&	3.638	&	0.299	&	23906.9	&	0.722	&	0.264	&	459.0	\\
ZTF	J080542.98-143036.3	&	1.507	&	0.287	&	3735.6	&	0.869	&	0.559	&	2916.7	\\
ZTF	J094826.35+253810.6	&	1.140	&	0.164	&	3282.9	&	0.696	&	0.149	&	6380.2	\\
ZTF	J102254.00-080327.3	&	2.561	&	0.397	&	2351.7	&	0.545	&	0.645	&	1867.7	\\
ZTF	J102653.47-101330.3	&	1.245	&	0.103	&	2643.1	&	0.859	&	0.778	&	4237.7	\\
ZTF	J104906.96-175530.7	&	1.380	&	0.194	&	2297.8	&	0.692	&	0.583	&	4344.9	\\
ZTF	J122009.98+082155.0	&	1.995	&	0.268	&	7057.0	&	0.628	&	0.558	&	2103.4	\\
ZTF	J125620.57+211725.8	&	1.346	&	0.102	&	3565.5	&	0.577	&	0.679	&	5920.2	\\
ZTF	J130228.34-003200.2	&	3.495	&	0.175	&	21428.7	&	0.684	&	0.387	&	1041.9	\\
ZTF	J134151.70-062613.9	&	1.783	&	0.122	&	1820.4	&	0.717	&	0.660	&	1660.7	\\
ZTF	J140036.65+081447.4	&	2.233	&	0.225	&	10978.5	&	0.738	&	0.542	&	1454.0	\\
ZTF	J140423.86+065557.7	&	3.020	&	0.398	&	6120.1	&	0.812	&	0.591	&	819.3	\\
ZTF	J140537.34+103919.0	&	0.944	&	0.096	&	2169.9	&	0.844	&	0.778	&	7951.8	\\
ZTF	J140702.57+211559.7	&	1.517	&	0.257	&	9139.6	&	0.845	&	0.433	&	4101.0	\\
ZTF	J162644.18-101854.3	&	1.127	&	0.206	&	1910.0	&	0.613	&	0.094	&	7043.6	\\
ZTF	J163421.00-271321.7	&	1.614	&	0.126	&	4008.4	&	0.605	&	0.382	&	3330.4	\\
ZTF	J164441.18+243428.2	&	1.375	&	0.096	&	2186.4	&	0.691	&	0.729	&	4482.0	\\
ZTF	J180256.45-005458.3	&	1.331	&	0.147	&	2572.3	&	0.810	&	0.612	&	4085.2	\\
ZTF	J195456.71+101937.5	&	1.434	&	0.443	&	2361.5	&	0.618	&	
0.129	&	3772.4	\\
\hline
        \end{tabular}
    \end{center}
\end{table*}

\begin{table*}[h!]
    \caption{Most probable solution (mode) for the stellar input parameters.}
    \label{tab:moda}
    \begin{center}
        \begin{tabular}{lccrccr}
            \hline
            Target & $M_{1,{\rm in}}$  & $M_{2,{\rm in}}$  & $P_{\rm in}$  & $e_{\rm in}$ & $\alpha_{\rm CE}$ & Age  \\
                             & (M$_{\odot}$) & (M$_{\odot}$) & (days) &  & & (Myr) \\
            \hline
            \hline
ZTF	J041016.82-083419.5	&	1.314	&	0.119	&	300	&	0.976	&	0.972	&	3900	\\
ZTF	J051902.06+092526.4	&	1.885	&	0.171	&	300	&	0.900	&	0.355	&	2250	\\
ZTF	J052848.24+215629.0	&	3.194	&	0.180	&	3900	&	0.825	&	0.526	&	1003	\\
ZTF	J053708.26-245014.6	&	1.482	&	0.198	&	600	&	0.975	&	0.990	&	2820	\\
ZTF	J061530.96+051041.8	&	2.193	&	0.514	&	500	&	0.980	&	0.335	&	1275	\\
ZTF	J063808.71+091027.4	&	2.695	&	0.404	&	1000	&	0.890	&	0.260	&	710	\\
ZTF	J063954.70+191958.0	&	2.850	&	0.206	&	600	&	0.920	&	0.550	&	702	\\
ZTF	J064242.41+131427.6	&	2.427	&	0.147	&	1700	&	0.915	&	0.640	&	965	\\
ZTF	J065103.70+145246.2	&	1.515	&	0.236	&	900	&	0.905	&	0.184	&	2180	\\
ZTF	J070458.08-020103.3	&	1.341	&	0.338	&	600	&	0.920	&	0.085	&	4650	\\
ZTF	J071759.04+113630.2	&	1.880	&	0.289	&	800	&	0.930	&	0.445	&	1560	\\
ZTF	J071843.68-085232.1	&	3.598	&	0.299	&	4200	&	0.918	&	0.315	&	460	\\
ZTF	J080542.98-143036.3	&	1.315	&	0.286	&	200	&	0.960	&	0.385	&	1470	\\
ZTF	J094826.35+253810.6	&	1.096	&	0.164	&	600	&	0.920	&	0.116	&	5010	\\
ZTF	J102254.00-080327.3	&	2.660	&	0.398	&	550	&	0.820	&	0.870	&	1730	\\
ZTF	J102653.47-101330.3	&	1.293	&	0.103	&	500	&	0.957	&	0.955	&	2950	\\
ZTF	J104906.96-175530.7	&	1.645	&	0.194	&	250	&	0.930	&	0.330	&	2300	\\
ZTF	J122009.98+082155.0	&	2.220	&	0.268	&	1600	&	0.885	&	0.558	&	1650	\\
ZTF	J125620.57+211725.8	&	1.455	&	0.102	&	900	&	0.905	&	0.835	&	4980	\\
ZTF	J130228.34-003200.2	&	3.514	&	0.174	&	2400	&	0.915	&	0.465	&	1038	\\
ZTF	J134151.70-062613.9	&	1.782	&	0.122	&	1100	&	0.685	&	0.770	&	1640	\\
ZTF	J140036.65+081447.4	&	2.295	&	0.225	&	400	&	0.955	&	0.510	&	1225	\\
ZTF	J140423.86+065557.7	&	2.910	&	0.399	&	8600	&	0.880	&	0.460	&	680	\\
ZTF	J140537.34+103919.0	&	0.912	&	0.096	&	200	&	0.925	&	0.695	&	6250	\\
ZTF	J140702.57+211559.7	&	1.810	&	0.257	&	200	&	0.980	&	0.240	&	2350	\\
ZTF	J162644.18-101854.3	&	1.140	&	0.206	&	1350	&	0.570	&	0.098	&	6500	\\
ZTF	J163421.00-271321.7	&	1.806	&	0.126	&	700	&	0.910	&	0.452	&	2480	\\
ZTF	J164441.18+243428.2	&	1.521	&	0.096	&	300	&	0.910	&	0.993	&	3000	\\
ZTF	J180256.45-005458.3	&	1.590	&	0.147	&	1350	&	0.924	&	0.440	&	2200	\\
ZTF	J195456.71+101937.5	&	1.368	&	0.444	&	1200	&	0.730	&	0.108	&	3900	\\
\hline
        \end{tabular}
    \end{center}
\end{table*}

\end{document}